\documentclass[%
 reprint,
 amsmath,amssymb,
 aps, 
]{revtex4-1}

\usepackage{graphicx}% Include figure files
\usepackage{dcolumn}% Align table columns on decimal point
\usepackage{bm}% bold math
\usepackage{xcolor}% color
\usepackage{xr} % cross-referencing between documents
\usepackage{subcaption}
\usepackage[justification=RaggedRight]{caption}
\usepackage{subfig}
\begin{document}

\title{Equilibrium phases and phase transitions in %a model of 
multicritical magnetic polymers}

\author{Alberto Raiola$^1$, Emanuele Locatelli$^1$, Davide Marenduzzo$^2$ and Enzo Orlandini$^1$ \\
$^1$  Dipartimento di Fisica e Astronomia e Sezione INFN, Universit\`a di Padova, \\ Via Marzolo 8, I-35131 Padova, Italy \\
$^2$ SUPA, School of Physics and Astronomy, University of Edinburgh, Peter Guthrie Tait Road, Edinburgh, EH9 3FD, UK \\}

\begin{abstract}
Magnetic polymers are examples of composite soft materials in which the competition between the large configurational entropy of the soft substrate (polymer) and the magnetic interaction may give rise to rich equilibrium phase diagrams as well as non-standard critical phenomena.
Here, we study a self-avoiding walk model decorated by Ising spins of value $0$ and $\pm 1$ that interact according to a Blume-Emery-Griffith-like Hamiltonian.
By using mean-field approximations and Monte Carlo simulations, we 
report the existence of three distinct equilibrium phases: swollen disordered, compact ordered, and compact disordered. Notably, these phases are separated by phase boundaries that meet at multicritical points, whose nature and location are tunable and depend on the strength of the interactions. In our conclusion, we discuss the relevance of the phase diagrams we have obtained to the physics of magnetic polymers and their application to chromatin biophysics.
\end{abstract}
\maketitle

\section{Introduction}
Models of magnetic polymers, where each monomer carries a magnetic moment or spin, are an interesting class of interacting systems that have recently received much attention in polymer physics, biophysics, and statistical mechanics. There are multiple reasons for this. %are several: (i) 
First, the flexibility of the polymeric substrate and its relatively low density are important features to exploit in organic magnetic materials with technological applications in, for instance, communication and information~\cite{rajca2001magnetic,miller2014organic,boudouris2024spin}. %; (ii) 
Second, from the statistical mechanics perspective, they are examples of interacting systems where the spatial organisation (entropy)  of the polymer chain and the magnetic interactions between spins may give rise to several conformational phases and phase transitions~\cite{garel1999phase,luo2006finite,papale2018ising,foster2021critical,rudra2023critical}. %; (iii) 
Finally, in the last few years, models of magnetic polymers have been successfully exploited to understand how the interplay between the chromatin folding in 3D and the epigenetic landscape in 1D, regulated by  histone marks, can contribute to shaping the genome organization in the nucleus~\cite{coli2019magnetic,michieletto2016polymer}.

A precursor of these models, introduced in~\cite{garel1999phase}, described the polymer as a self-avoiding walk (SAW) on a hypercubic lattice whose monomers are decorated by Ising spin variables $\sigma_i=\pm 1$ that interact via the standard ferromagnetic Ising Hamiltonian among themselves, when they are nearest neighbours \emph{in 3D space}, and with an external magnetic field $h$.  By performing both a mean-field analysis and Monte Carlo simulations of the 3D system (the SAW on the cubic lattice) it was shown that, by increasing the ferromagnetic coupling,  this model displays, at $h=0$, a first-order phase transition between a swollen and paramagnetic (disordered) phase (SD) and a compact ferromagnetic (ordered) phase (CO).  The first-order character is a notable feature of this magnetic model, since in a standard (i.e., non-magnetic) polymer collapse the ($\Theta$) phase transition is known to be second order~\cite{lifshitz1978some}.
Later studies have either focused on better determining the location and nature of the transition in $D=2,3$ by numerical simulations and finite-size scaling~\cite{luo2006finite,foster2021critical}, or studied the emergence of a ferromagnetic phase transition as a function of the fractal dimension of the polymer substrate~\cite{papale2018ising}.

Recently, a model of magnetic polymers with 3-state variables ($q_i=1,-1,0$) each %describing 
representing an epigenetic mark, has been introduced to describe the interplay between the 3D organisation of the eukaryotic genome and the 1D epigenetic profile of biochemical marks on chromatin (the DNA-histone composite polymer which provides the building block of chromosomes in eukaryotes)~\cite{coli2019magnetic}. Analytical calculations based on mean-field approximation and molecular dynamics simulations have shown that a first-order phase transition rules both the spreading of a single epigenetic mark and the folding of the chromosome into a compact structure. The first-order nature of the transition is important, as it endows the system with memory, which allows a cell to ``remember'' its state following cell division~\cite{michieletto2016polymer}.

Note that, similarly to the works cited above, this biophysical model can display only two equilibrium phases: a swollen disordered (SD) phase and a compact (magnetically) ordered (CO) phase. To observe a third phase corresponding to a compact chromatin phase where the epigenetic marks (spins) are incoherently distributed, as in gene deserts regions observed in \emph{Drosophila}~\cite{filion2010systematic}, an extension to a non-equilibrium system was needed\cite{michieletto2019nonequilibrium}.

An interesting problem that emerges from the previous studies is whether there exist spin models embedded in fluctuating filaments that can sustain more than two phases (SD and CO) and multiple phase transitions {\it at equilibrium}.  A first attempt to investigate this issue was made very recently in Ref.~\cite{nakanishi2024emergence} where, by introducing an energy offset $\epsilon_{off}$ in the interaction energy of a polymer Potts model, a  compact disordered (CD) phase was obtained above a critical value of $\epsilon_{off}$.

In this paper, we study how a model of a magnetic system with a rich phase diagram and multicritical points, once embedded in a polymer chain, can affect the polymer's magnetic and spatial organisation. This investigation is carried out analytically, at the mean-field level, and numerically, by simulating the magnetic polymer on a cubic lattice.

The paper is organized as follows. In section II we introduce the magnetic polymer model and derive the mean-field free energy density and the relevant order parameters such as the monomer density and the magnetization. In Section III the mean-field conformational/magnetic phase diagram and the nature of the several phase transitions found are determined as a function of the magnetic coupling parameters. % of the magnetic system. 
Section IV presents the  Monte Carlo simulations of a self-avoiding walk model in the cubic lattice and compares the obtained numerical results with the mean-field findings along some relevant lines of the phase diagram. Finally, Section V is devoted to discussions and conclusions.

\section{Model and mean-field free energy density}
 \begin{figure}
    \centering    \includegraphics[width=\linewidth]{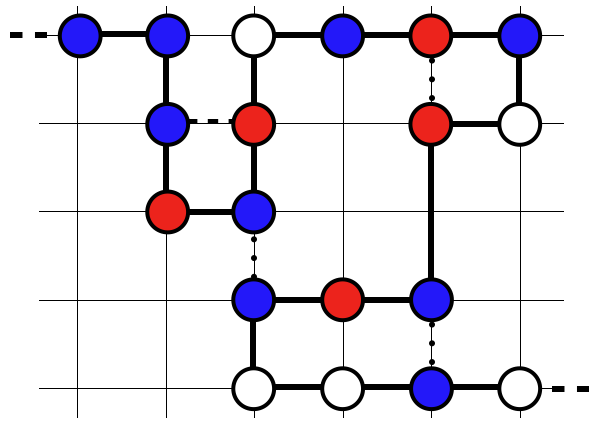}
    \caption{Cartoon of a self avoiding walk on a two dimensional square lattice. The thick black line represents the polymer backbone, the thin straight lines represent the underlying  lattice. Each vertices $i$ of the walk holds a spin variable $S_i$ which can be equal to $-1$ (red circle), $0$ (white circle) or $+1$ (blue circle). A dotted line is drawn in correspondence of pairs of spin variables with $S_i=\pm 1$ which are not nearest neighbors on the walk but in 3D space.} %; the dashed line is used in the same situation when two opposite sign spin variables interact. }
    \label{Cartoon}
\end{figure}
We consider the ensemble of the $N$-steps self-avoiding walks (SAWs) $\{ \gamma \}$ on a lattice of coordination number $z$. By assigning the set $\{S_i\}$ of spin variables  $S_i \in \{0, \pm 1\}$ to the lattice sites occupied by a SAW, $\gamma$, the Hamiltonian of the system we wish to study is given by
\begin{eqnarray}
    \mathcal H(\gamma,\{S_i\}) &=& -\frac{J}{2}\sum_{i,j}S_i \Lambda_{i,j}^{\gamma}S_j - \frac{K}{2} \sum_{i,j}S_i^2 \Lambda_{i,j}^{\gamma}S_j^2 \nonumber \\&-& \Delta \sum_{i = 1}^N \left (1-S_i^2 \right).
    \label{eq_Hamiltonian}
\end{eqnarray}

The first term in Eq.~\eqref{eq_Hamiltonian} is the standard ferromagnetic interaction with $J>0$ being the exchange energy. The interaction is restricted to the pairs $(i,j)$ of nearest neighbor (NN) sites of the lattice occupied by the SAW $\gamma$ -- this is achieved via the adjacency matrix $\Lambda_{i,j}^{\gamma}$, which is defined as  $\Lambda_{i,j}^{\gamma} = 1$ if $i,j$ are NN on the lattice and $\Lambda_{i,j}^{\gamma} = 0$ otherwise.  The second term, characterized by the parameter $K$, provides an energy gain for any pair of non-zero NN spins ($S_i=\pm 1$), irrespective of their sign. Finally, the last term weights the neutral sites on the walk via the dilution parameter $\Delta$ ($S_i=0$). This term introduces a ``mitigation'' of the ferromagnetic interaction whose amount (i.e., fraction of neutral spins) is ruled by $\Delta$; thus, we can interpret $\Delta$ as a chemical potential. Note that the Hamiltonian Eq.~\eqref{eq_Hamiltonian} was studied by Blume, Emery, and Griffiths (BEG) on a lattice, as a spin-1 model of a diluted uniaxial ferromagnet or of a $\mathrm{He^3}-\mathrm{He^4}$ mixture~\cite{BEG}. The key emergent feature of the BEG model is the presence of a tricritical point, separating a line of second-order (continuous) phase transitions from a line of first-order (discontinuous) transitions. In our model, we shall distinguish tricritical points, defined in such a way, from multicritical points, where multiple transition lines (of different order) meet, and triple points, which mark points at which three different phases are in equilibrium. 

By taking the Boltzmann factor associated to Eq.~\eqref{eq_Hamiltonian} and summing over the spins and the SAW configurations $\{\{S_i\},\{\gamma\}\}$ we get the partition function of the model:
\begin{widetext}
\begin{equation}
  \mathcal Z = \sum_{\gamma \in \{\gamma\}} \sum_{\{ S \} } \exp \left ( \frac{\beta J}{2} \sum_{i,j}S_i \Lambda_{i,j}^{\gamma} S_j  +  
    \frac{\beta K}{2} \sum_{i,j}S_i^2 \Lambda_{i,j}^{\gamma} S_j^2 + \beta \Delta \sum_{i = 1}^N \left (1- S_i^2 \right)\right ).  
    \label{partition}
\end{equation}
\end{widetext}

A mean-field estimate for the free energy density corresponding to Eq.~\eqref{partition} can be obtained by first decoupling the quadratic and bi-quadratic terms via a double Hubbard-Stratonovich transformation involving two local fields $\{\phi_i\}$ and $\{\alpha_i\}$, and then performing the sum over the spin degrees of freedom and the SAW configurations  (see Appendix A). This procedure gives as a result
\begin{widetext}
\begin{equation}
f(\beta, \phi, \alpha, \rho) = -\frac{1}{\beta N} \log (\mathcal{Z}) = -\frac{1}{\beta}\log \frac{z}{e} + \frac{1 - \rho}{\beta \rho} \log (1 - \rho) + \frac{1}{2\beta^2 z \rho } \left (\frac{\phi^2}{J} + \frac{\alpha^2}{K}  \right ) - \frac{1}{\beta} \log (1 + 2e^{-\beta \Delta + \alpha} \cosh (\phi)) - \Delta 
\label{fed},
\end{equation}
\end{widetext}
where $\log$ denotes the natural logarithm. The free energy density Eq.~\eqref{fed} allows to obtain the magnetization per spin $m = \langle S_i \rangle$ and the average concentration of neutral spins $x$ as:
\begin{equation}
m = \langle S_i \rangle =  \frac{2e^{-\beta \Delta + \alpha}\sinh \phi}{1 + 2e^{-\beta \Delta + \alpha}\cosh \phi},
\label{mformula}
\end{equation}
\begin{equation}
x = 1 - \langle S_i^2 \rangle =  \frac{1}{1 + 2 e^{-\beta \Delta + \alpha}\cosh \phi} ,
\label{xformula}
\end{equation}
respectively. Minimizing Eq.~\eqref{fed} with respect to $\phi$, $\alpha$, and $\rho$ we obtain 
\begin{align}
\frac{\phi}{\beta J z \rho} - \frac{\sinh(\phi)}{\frac{e^{\beta \Delta - \alpha}}{2} + \cosh(\phi )} &= 0  \label{kmfe1}\\
\frac{\alpha}{\beta K z \rho} - \frac{\cosh(\phi)}{\frac{e^{\beta \Delta - \alpha}}{2} + \cosh(\phi )} &= 0 \label{kmfe2}\\
\rho + \log(1 - \rho) + \frac{1}{2\beta z} \left ( \frac{\phi^2}{J} + \frac{\alpha^2}{K} \right ) &= 0 \label{kmfe3}.
\end{align}
Comparing Eq.~\eqref{mformula} and ~\eqref{kmfe1} we the self-consistent condition $\phi = \beta J z \rho m$.
Similarly, matching  Eqs.~\eqref{xformula} and~\eqref{kmfe2} we get $\alpha = \beta K z \rho (1 - x)$.

\begin{figure*}[t]
    \centering
\includegraphics[width=\linewidth]{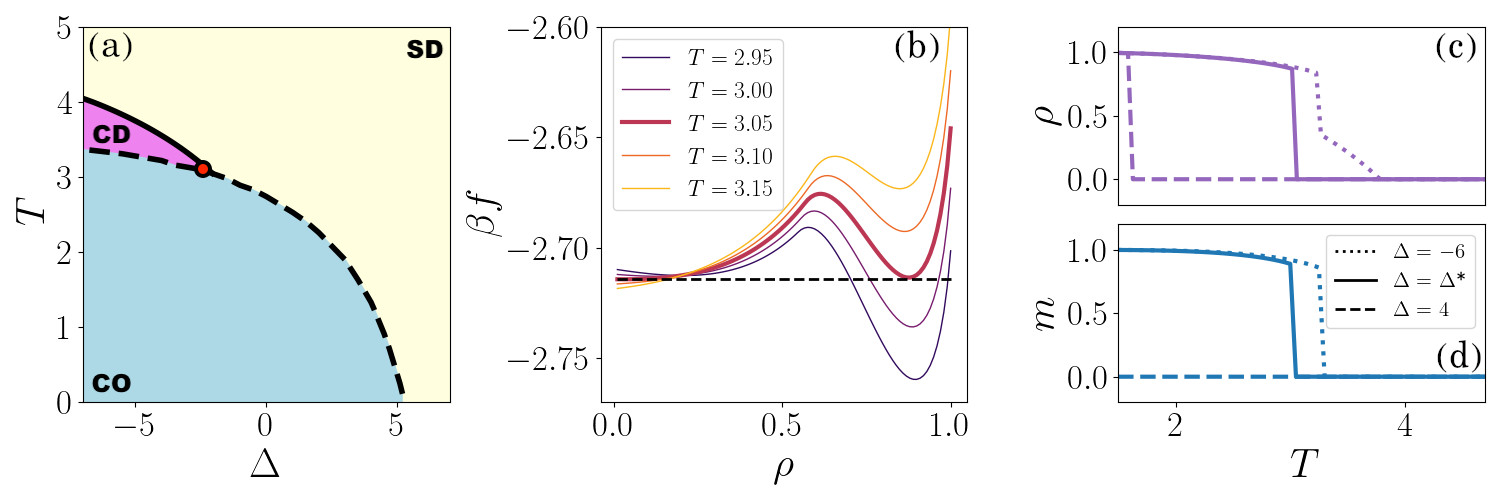}
\caption{ Mean-field results at $K/J=0.8$. 
(a) Equilibrium phase diagram in the $(\Delta,T)$ plane: note the emergence of three phases -- SD, CO and CD -- for  $\Delta < \Delta^* \simeq -2.04$. 
Dashed (first order) and solid (continuous) lines show the corresponding phase boundaries. The location of the multicritical point (here coinciding with the triple point) $(\Delta^*, T^*) \simeq (-2.04, 3.05)  $ is highlighted with a red circle. (b) $\rho$ dependence of the free energy density at $\Delta = \Delta^*$. The thickest red curve marks the case $T = T^*$. % and displays two equal height minima.
(c-d) Temperature dependence of  the polymer density $\rho$ (c) and the average magnetization per spin $m$ (d) at three different values of $\Delta$. 
}
    \label{c1c2}
\end{figure*}

\section{Mean-field phase diagrams}

To analyse the mean-field free energy Eq.~\eqref{fed}  in the $(\Delta,T)$ plane 
we first divided Eq.~\eqref{kmfe2} by Eq.~\eqref{kmfe1}, obtaining
\begin{equation}
\alpha(\phi) =\frac{K}{J} \phi \coth \phi.
\label{alpha_vs_phi}
\end{equation}
Following the steps reported in subsection B.3 of the Appendix, we next obtain a virial expansion of  Eq.~\eqref{fed}, up to order $\phi^6$. We then numerically estimate the signs of $a(\beta,\Delta)$, $b(\beta,\Delta)$, and $c(\beta,\Delta)$ as functions of $T=1/\beta$ ($k_B=1$) and $\Delta$ (see Figure~\ref{fig:abc_CDCO} of the Appendix). Since we are interested in the competition between the ferromagnetic strength, $J$, and the self-attraction strength, $K$, we fix $J=1$ and consider four cases: $K = 0.8, 1.8, 2.3, 3.0$. For each case, we discuss the phase diagram and the nature of the critical points and transition lines by looking at the free energy as a function of $\rho$ as well as $m$ and $\rho$ as a function of $T$. 

\subsubsection{The $K/J < 1$ case}
We start from the mean-field phase diagram in the $(\Delta,T)$ plane, reported in Figure~\ref{c1c2}a, for the case $K/J=0.8$, which is representative of the mean-field phase behaviour when $K/J<1$.

The first significant feature of note is the presence of three equilibrium phases: (i) a magnetically disordered phase, where the polymer chain is extended (\emph{swollen disordered or SD} phase); (ii) a \emph{compact ordered or CO} phase where the polymeric substrate is globular and magnetically ordered (i.e. with most of the spins either up or down); (iii) a magnetically disordered but compact phase (\emph{compact disordered or CD} phase) where the polymer density is non-zero but the spins are in the paramagnetic phase. The CD phase is especially notable here, because it is generally difficult to be observed at equilibrium in simpler models of magnetic polymers~\cite{michieletto2019nonequilibrium,nakanishi2024emergence}. 

The phase diagram shows the presence of three phase boundary curves: (i) two first-order phase transition lines, one between the CO and SD phases and the other between the CO and CD phases, and (ii) a second-order (continuous) phase transition between the CD and the SD phase. The three boundaries meet at a multicritical point, at ($(\Delta,T)=(\Delta^{*},T^{*})$ (the red circle in Fig.~\ref{c1c2}a): in the present case of $K/J=0.8$, this is also a triple point as the CO, CD and SD phases are all in equilibrium there.

The nature of the multicritical point is elucidated by the $\rho$ dependence of the free energy density, estimated at $\Delta=\Delta^*$ (see Figure~\ref{c1c2}b).  For $T>T^*$ the absolute minimum of the free energy is at $\rho=0$ (Swollen phase, SD), while in the region $T<T^*$ the global minimum is located at $\rho \approx 0.9$ (corresponding to the Compact ordered phase, CO); at $T=T^*$ (the thickest red curve) two equal height minima are observed, one at $\rho=0$ and one at $\rho \simeq 0.85$. Additionally, as we cross the multicritical point by varying $T$ at fixed $\Delta=\Delta^* \simeq -2.04$ (see solid lines in Fig.~\ref{c1c2}c,d), both the $m$ vs $T$ and the $\rho$ vs $T$ curves display a finite jump.

Examining the behavior of the curve $\rho(T)$ at different values of $\Delta$ clarifies the nature of the transition lines. At $\Delta=-6$ (see the dotted curve in Figure~\ref{c1c2}c), upon increasing $T$, we encounter a distinct discontinuity near the CO/CD boundary; after this jump, the curve gradually decreases, approaching zero as it approaches the CD/SD phase boundary.
This suggests a first-order transition at the CO/CD boundary (confirmed by the finite discontinuity in the $m$ vs $T$ curve, Figure~\ref{c1c2}c), followed by a continuous phase transition at the CD/SD boundary.

Finally, the finite discontinuities observed in the $m$ vs $T$ and $\rho$ vs $T$ curves at $\Delta = 4$ (see dashed curves in Figure~\ref{c1c2}c,d) confirm the first-order nature of the CD/SD phase transition.

\begin{figure*}[t]
    \centering
\includegraphics[width=\linewidth]{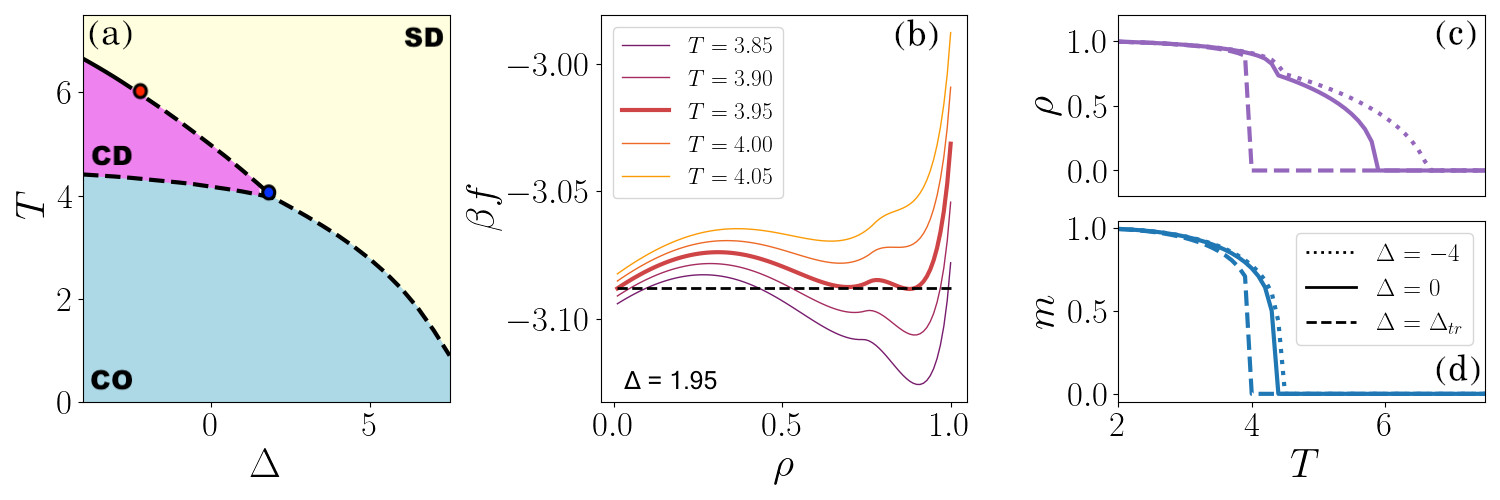}

\caption{{Mean-field results at $K/J = 1.8$.} (a) Equilibrium phase diagram in the $(\Delta, T)$.  The SD/CD phase boundary is partitioned into a continuous (solid) and first-order transition curve (dashed curve). The corresponding tricritical point, located at $(\Delta^*, T^*) \approx (-2.46 ,6.08)$ is highlighted by a red circle, while the blue circle marks the triple point $(\Delta_{\rm tr}, T_{\rm tr}) \approx (1.95,3.95)$. (b): Free energy density landscapes at $\Delta = \Delta_{\rm tr} = 1.95$ as a function of the polymer density $\rho$. The coexistence of the three phases at the triple point is marked by the three, equal height, minima at $T_{\rm tr} = 3.95$ (thickest line). (c-d) Density $\rho$ (c) and magnetization $m$ (d) as functions of $T$ at  $\Delta = -4$ (dotted line), $\Delta = \Delta_{\rm tr}$ (solid line) and $\Delta = 4$ (dot-dashed line).}
    \label{d1d2}
\end{figure*}

\subsubsection{The $K/J > 1$ case}

When the self-attraction strength exceeds the ferromagnetic interaction, the phenomenology becomes more complex.

We start from $K/J=1.8$, where the analysis of the corresponding virial coefficients (as shown in Fig.~\ref{fig:abc_CDCO}c) reveals two key properties of the system:
(i) the compact disordered (CD) phase region expands towards higher $\Delta$ values, indicating stability at greater dilutions;
(ii) as we progress from lower dilution regimes (i.e., from negative $\Delta$ values), the nature of the CD/SD phase boundary changes from continuous to first-order.
This latter switch coincides with a sign change in the third virial coefficient along this curve, revealing the presence of a tricritical point at $\Delta^{*} = -2.46$ and $T^{*} = 6.08$. Simultaneously, the triple point shifts to $\Delta_{\rm tr} = 1.95$, $T_{\rm tr} = 3.95$. In other words, the topology of the phase diagram has fundamentally changed, as the multicritical point at $K/J=0.8$ has split into a tricritical and a triple point (see  Fig.~\ref{d1d2}a).

The three-phase coexistence at the triple point can be understood by looking at the $\rho$ dependence of the free energy density computed at $\Delta_{\rm tr}$ and for some values of $T$  above, below and at $T_{tr}$ (see Figure~\ref{d1d2}b). There are three key points to note. First, %(i) 
for $T<T_{\rm tr}$ there is a unique absolute minimum located at  $\rho \simeq 0.88$, that corresponds to the CO phase. %(ii)
Second, for $T>T_{tr}$ the absolute minimum is at $\rho=0$, as expected in the swollen phase. %; (iii) 
Third, at the triple point temperature $T=T_{\rm tr}$, the system exhibits three minima of equal depth in its free energy landscape. One of these minima occurs at $\rho=0$, while the other two occur at $\rho=0.71$ and $\rho=0.88$ respectively. This free energy profile describes the coexistence of three distinct phases, two compact (ordered and disordered) and one swollen (disordered). The triple minimum was absent in the lower $K/J$ case, as the transitions between the phases were not all first-order.

This scenario is confirmed in Figs.~\ref{d1d2}c-d where we report the temperature behavior of $\rho$ and $m$ for $\Delta = -4.0$ and $\Delta_{\rm tr}$.
\begin{figure*}[t]
    \centering
\includegraphics[width=\linewidth]{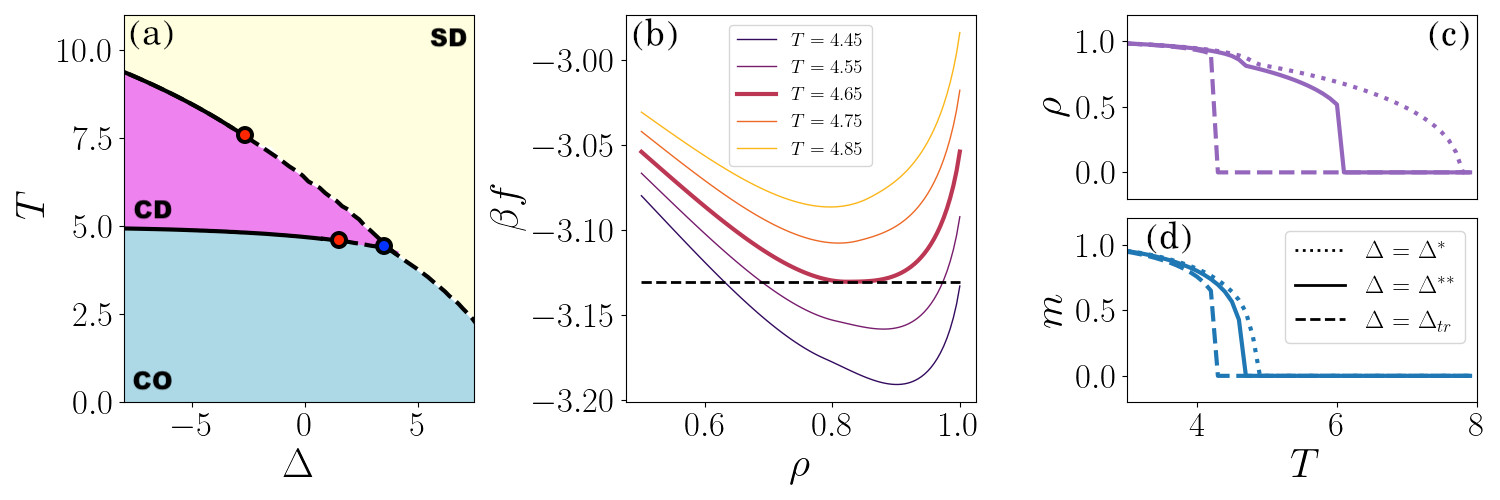}
    \caption{{Mean-field results at $K/J = 2.3$} (a) Equilibrium phase diagram in the $(\Delta, T)$ plane. The two red circles highlight the two multicritical points located at ($\Delta^*$, $T^*$) = ($-3.15, 7.76 $) and ($\Delta^{**}$, $T^{**}$) = ($0.63 , 4.65$). The blue circle  locates the triple point ($\Delta_{\rm tr}$, $T_{\rm tr}$) = ($3.98, 4.23$) (b) Free energy profiles for $\Delta = \Delta^{**} = 0.63$. The extended flatness of the thicker red line suggests the tricritical nature of the $(\Delta^{**}, T^{**})$ point. (c-d) Temperature dependence of the polymer density $\rho$ (c) and magnetization $m$  (d) for $\Delta = \Delta^*$ (i.e. crossing the leftmost multicritical point), $\Delta = \Delta^{**}$ (crossing the second multicritical point) and $\Delta = \Delta_{\rm tr}$ (crossing the triple point). % at the confluence of the three phases).
    }
    \label{e1e2}
\end{figure*}

For yet larger values of $K/J$, the CD  phase expands further into the $\Delta > 0$ region and, notably, for $K/J \simeq 1.805$ the nature of the CO/CD phase boundary splits into a second-order and first-order transition curves that are governed by a novel critical point.
Such a phenomenon is evidenced by the analysis of the virial coefficients reported in Fig.~\ref{fig:abc_CDCO}c,d, which also shows that this crossover is indeed a new, second, tricritical point ($\Delta^{**}$, $T^{**}$). For $K/J \simeq 1.805$, such a point is located at $\Delta^{**} = -\infty$ and moves to larger values of $\Delta$ with increasing $K/J$. As showcased in the phase diagram of Figure~\ref{e1e2}a, for $K/J =2.3$ the point is located at $\Delta^{**}=0.63, T^{**} =4.65$. The corresponding free energy profile is illustrated by the thickest curve of Figure~\ref{e1e2}b and also by the $m$ vs $T$ and $\rho$ vs $T$ curves in Figures~\ref{e1e2}c-d.

\begin{figure*}[t]
    \centering
\includegraphics[width=\linewidth]{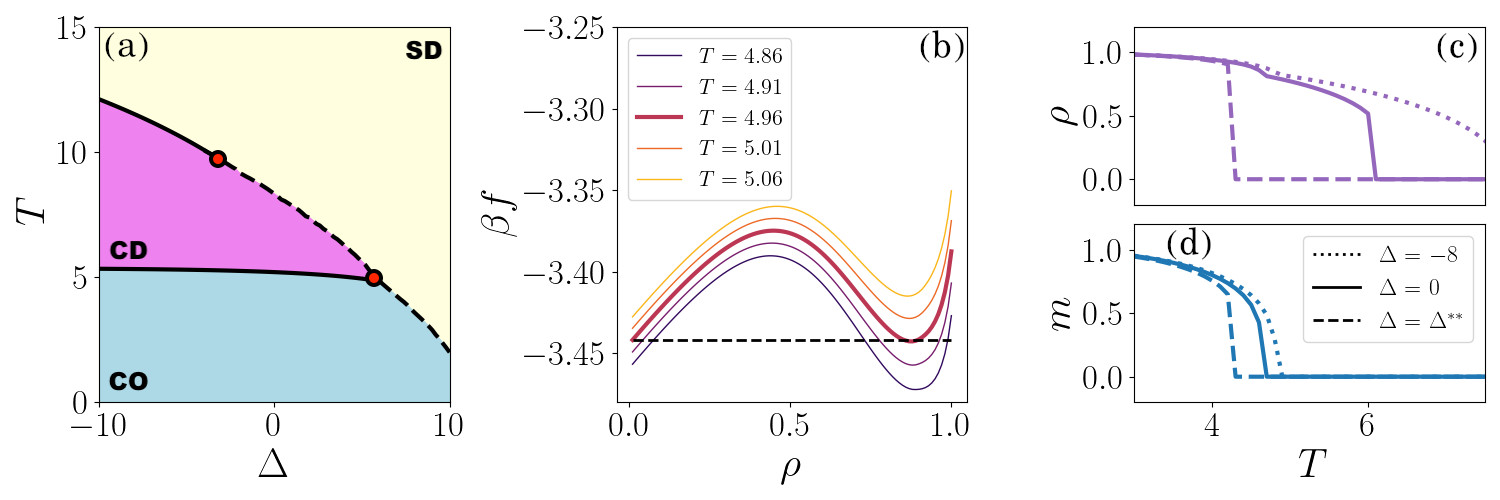}
    \caption{{Mean-field results at $K/J = 3.0$} (a) Equilibrium phase diagram in the $(\Delta, T)$ plane. The two red circles highlight the location of the two multicritical points, ($\Delta^*$, $T^*$) = ($-4.10, 10.13 $) and ($\Delta^{**}$, $T^{**}$) = ($6.00, 4.83$) (b) Density dependence of the free energy at $\Delta = \Delta^{**}$. The two equal-height minima displayed by the red thickest curve signal the coexistence of a swollen $\rho = 0$ and a compact $\rho = 0.87$ phase. (c - d) Temperature dependence of the polymer's density $\rho$ (c) and magnetization $m$ (d) for $\Delta = -8$ (i.e. crossing the two continuous phase boundaries), $\Delta = 0$ (crossing the continuous CO/CD boundary and the first-order CD/SD boundary) and $\Delta = \Delta^{*}$ (crossing the multicritical point). %at the confluence of the three phases.)
    }
    \label{e1e2K3}
\end{figure*}

A consequence of the phenomenology discussed above is that, when $K/J \gg 1$, the entire CO/CD phase boundary becomes continuous. Consequently, the original triple point (indicated by a blue circle in Figure~\ref{e1e2}a)  
is superseded by the multicritical point at $(\Delta^{**}, T^{**})$ as in the phase diagram of Fig.~\ref{e1e2K3}a, obtained for $K/J=3$. At this point, two first-order phase boundaries converge with a continuous one. The corresponding free energy profile is reported as the thicker line in Figure~\ref{e1e2K3}b: one can notice the concomitant presence of one minimum at $\rho=0$ (swollen) and one higher-order minimum located at $\rho=0.86$. The $T$ dependence of $m$ and $\rho$ are reported in Figure~\ref{e1e2K3}c-d for three special cases, as we cross (i) the two continuous phase boundaries ($\Delta=-6$), (ii) the continuous and first-order phase boundaries ($\Delta=0$) and (iii) the multicritical point.

This concludes our mean-field analysis. In the next Section, we ask whether the rich scenario obtained within the mean-field theory is confirmed by the numerical results of a 3D lattice model magnetic polymer simulated using Monte Carlo methods.

\section{Numerical results}
\begin{figure*}[t]
    \centering
    \includegraphics[width=.3\linewidth]{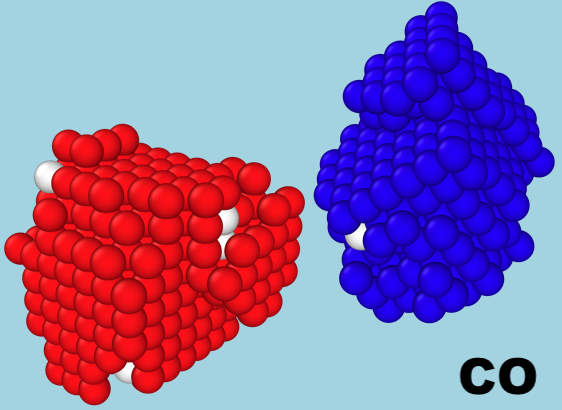}
     \includegraphics[width=.3\linewidth]{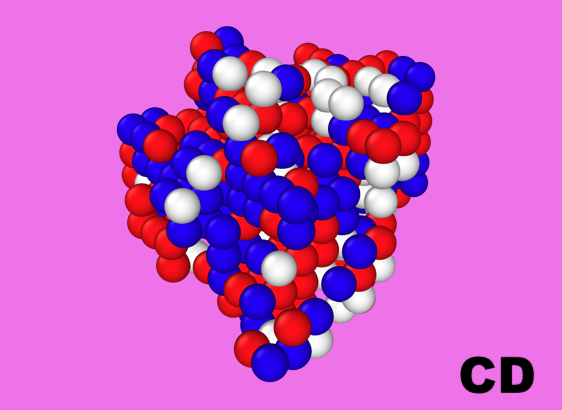}
    \includegraphics[width=.3\linewidth]{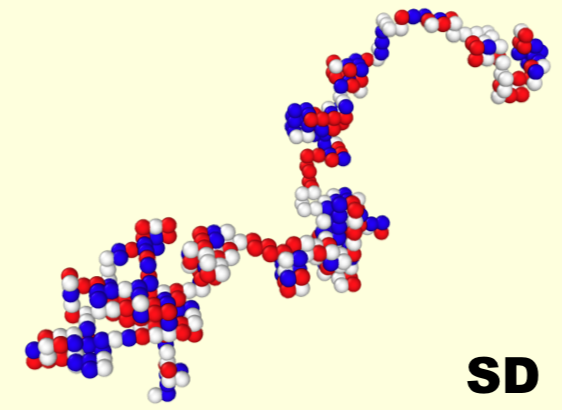}
    \caption{Snapshots of typical MC configurations of a magnetic self-avoiding walk with $N=400$, $K/J=3$ and $\Delta=-5$. The three panels were sampled respectively in the CO (left, $T=2.2$), CD (middle, $T=4.5$), and SD (right, $T=10$)  equilibrium phases. Red, white, and blue monomers correspond to spin values  $S = -1,0$ and $1$, respectively. Note that in the CO phase, the polymer is greatly packed and almost all the monomers carry the same spin value; because of the $\mathbb Z_2$ symmetry, the two values of the magnetization $ m =\pm 1$ can occur with equal probability. }
    \label{Configurations}
\end{figure*}

To simulate the model defined by Eq.~\eqref{partition} in 3D space, we consider the set of self-avoiding walks on a cubic lattice where each vertex is decorated by a spin variable $S_i$, which can take on the values $0, \pm 1$.  The set of configurations $\{ \gamma, \{S\}\}$ is sampled by a Monte Carlo (MC) algorithm based on a collection of elementary moves. These are: (i) pivot moves, essential for the ergodicity of the algorithm~\cite{madras1990monte}; (ii) a set of local Verdier-Stockmayer-style moves~\cite{verdier1962monte} that are known to increase the mobility of the Markov chain in proximity of compact phase~\cite{Tesi}; (iii) a spin flip or Glauber dynamics to update the spin configurations along the walk.  For an $N$-steps walk a MC step (sometimes called sweep) consists in attempting $1$  pivot moves intercalated by $N/4$ local moves and $N$ spin updates. By using a Metropolis heat bath sampling with $\beta \equiv k_BT$, the resulting Markov chain at temperature $T$ is expected to converge to the equilibrium distribution
\begin{equation}
    \pi(\beta) = \frac{1}{Z(\beta)} \exp{\left (\,-\beta{\cal H}(\gamma,\{S\}) \, \right)},
\end{equation}
where $Z(\beta)$ and ${\cal H}(\gamma,\{S\})$ are given by Eq.~\eqref{eq_Hamiltonian} and Eq.~\eqref{partition} respectively. 

To enhance the sampling efficiency, we implement a multiple Markov chain algorithm. This approach involves running $N_p$ parallel Markov chains, each at a distinct, fixed temperature $T$.
A coupling between the Markov chains is established by trying to swap the configurations between chains which are contiguous in temperature space. By using a suitable swap protocol (here we have attempted a swap for each $5 \cdot 10^3$ MC step) one can show that the collection of parallel Markov chains is itself a Markov chain whose stationary distribution is the product of the Boltzmann distribution along each chain (see references~\cite{Tesi}). 

For each value of the parameter $K$  considered (at fixed $J=1$), we sample every $M=50$ attempted pivot moves and along  $N_p \approx 30$ parallel chains for systems of size $N\in \{ 50,100,200,300,400\}$.  This amounts to a total of at least $2 \cdot 10^7$ sampled configurations for each value of the set of parameters $(N,K,J)$. 

For each multiple Markov chain run we measure observables describing the state of the system. To characterise the polymer conformation, we compute the average number of contacts $\langle c\rangle$ -- i.e., the number of pairs of  nearest-neighbor lattice vertices occupied by the polymer that are not contiguous along the polymer backbone --, its variance per monomer $Var(c)\equiv \left (\langle n_{c}^{2} \rangle - \langle n_{c} \rangle^2 \right )/N$  and  the mean squared  radius of gyration 
\begin{equation}
R_{G}^{2} =  \frac{1}{N} \sum_{i = 1}^{N} \|\mathbf R_{i} - \mathbf R_{CM} \|^{2}, 
\end{equation}
where $\mathbf R_{CM} =\frac{1}{N} \sum_{i = 1}^{N} \mathbf R_{i}$ is the position of the centre of mass and $\mathbf R_{i}$ is the position of the $i$-th monomer.
To monitor the magnetic properties of the system, we estimate the degree of dilution, i.e., the concentration of neutral spins $x = 1- \langle S^2\rangle$, the \emph{magnetization per spin} $\langle m \rangle =  \frac{1}{N} \sum_i \langle S_i\rangle$, and the \emph{magnetic susceptibility per spin}
$\chi_{M} = \frac{\langle m^{2} \rangle - \langle m \rangle^{2}}{k_{B} T}$.
Finally, we also computed the average total energy $\langle E \rangle$ and the specific heat 
$C = \frac{\langle E^2 \rangle - \langle E \rangle^2}{k_B T^2}$.\\
To locate the phase boundaries, we look at the $T$ dependence of the specific heat, the magnetic susceptibility, and the variance of the number of contacts.
We further perform a finite size scaling (FSS) analysis on the specific heat, the magnetic susceptibility, the variance of the contacts, and the radius of gyration. FSS theory predicts, for instance, that the location of the peaks of the specific heat, $T^*(N)$,  shifts towards the critical temperature $T_c$ as $N\to \infty$, as 
\begin{equation}
    T^*(N) \sim T_c + A N^{-\frac{1}{2-\alpha}}.
\end{equation}
For a continuous phase transition such as, for instance, the polymer $\Theta$-point transition~\cite{vanderzande1998}, we know that $\alpha=0$. Hence, by plotting $T^*(N) \hbox{ vs } 1/\sqrt{N}$, a linear convergence toward the critical temperature is expected. On the contrary, a first-order transition corresponds to the value $\alpha=1$, which implies  that $T^*(N)$ vs $1/{N}$ should display a linear behaviour.\\ 
Similarly, we plot the value of the peaks of $C$, $\chi_M$, and $Var(c)/N$. FSS predicts that these peaks behave in the proximity of the critical region (i.e. for $N$ large enough) as 
\begin{equation}
H_{max} \sim N^{\frac{\alpha}{2-\alpha}},
\end{equation}
where again $\alpha=1$ for a discontinuous phase transition and $\alpha=0$ for a continuous one. In the latter case, a logarithmic term should dominate, and thus, by plotting the values of these peaks as a function of $\log{N}$, a linear growth is expected. \\
In addition, we estimate the order of the transitions by computing the fourth-order Binder cumulant of the distribution function of a given observable $O$: $B_n(O) = 1 -\langle O^4\rangle / 3\langle O^2\rangle^2$. According to FSS theory, the location of the minimum of $B_n(O)$ may be extrapolated to estimate the transition point. Importantly, if the phase transition is continuous, the values of $B_n(O)$ at the minimum should asymptotically approach $2/3$ with increasing $N$. In contrast, it should go to another limit if the transition is first-order~\cite{Bindercum}.

We estimate the errors associated with each observable $O$ by using the corresponding integrated autocorrelation times $\tau_O$. Some estimates of $\tau_O$ (given in units of the sampling, i.e., each $M=50$  attempted pivot moves for $O=E,m,R_G$) are reported in Appendix B for $\Delta=-5,0$ and $6$, respectively (see Table I, II, and III): as expected, sampled configurations are increasingly more correlated as $T$ is decreased. At the lowest values of $T$, the simulations performed yield only a few hundreds of fully uncorrelated configurations. This impact on the size of the error bars that increase rapidly as $T$ decreases, as exemplified in Figure~\ref{barreErrore} in the Appendix. % , where we report some observables with the corresponding error bars. 
To show the trends of the various observables more clearly, the error bars have been omitted from the other figures presented.

Although simulating polymers on the lattice is more efficient computationally with respect to off lattice models, a full numerical exploration of the $(\Delta, T)$ plane, even for a single specific $K/J$ value, is still virtually impossible.
We therefore restrict ourselves to the case $K/J=3$, where the mean-field approximation suggests the emergence of a rich phase diagram with multiple phase transitions and critical points.

In particular, we focus on the CO/CD and CD/SD phase transitions at different values of $\Delta$ ($\Delta=-5,0,6$). 

\subsection{Simulations for $K/J=3, \Delta=-5$} 
For negative values of $\Delta$, neutral spins are not favourable and the dilution is weak; here mean-field theory predicts the presence of two continuous phase transitions (see Fig.~\ref{e1e2K3}a). 
In Figure~\ref{Configurations} we report snapshots of typical configurations at $T = 2.2$, $T=4.5$, and $T=10$, where we expect the system to be, at equilibrium, in the CO, CD, and SD phases, respectively. The snapshots confirm the presence of all three phases, and importantly of the CD phase for the intermediate $T=4.5$ temperature. 

\begin{figure*}
    \centering
    \includegraphics[width=0.9\linewidth]{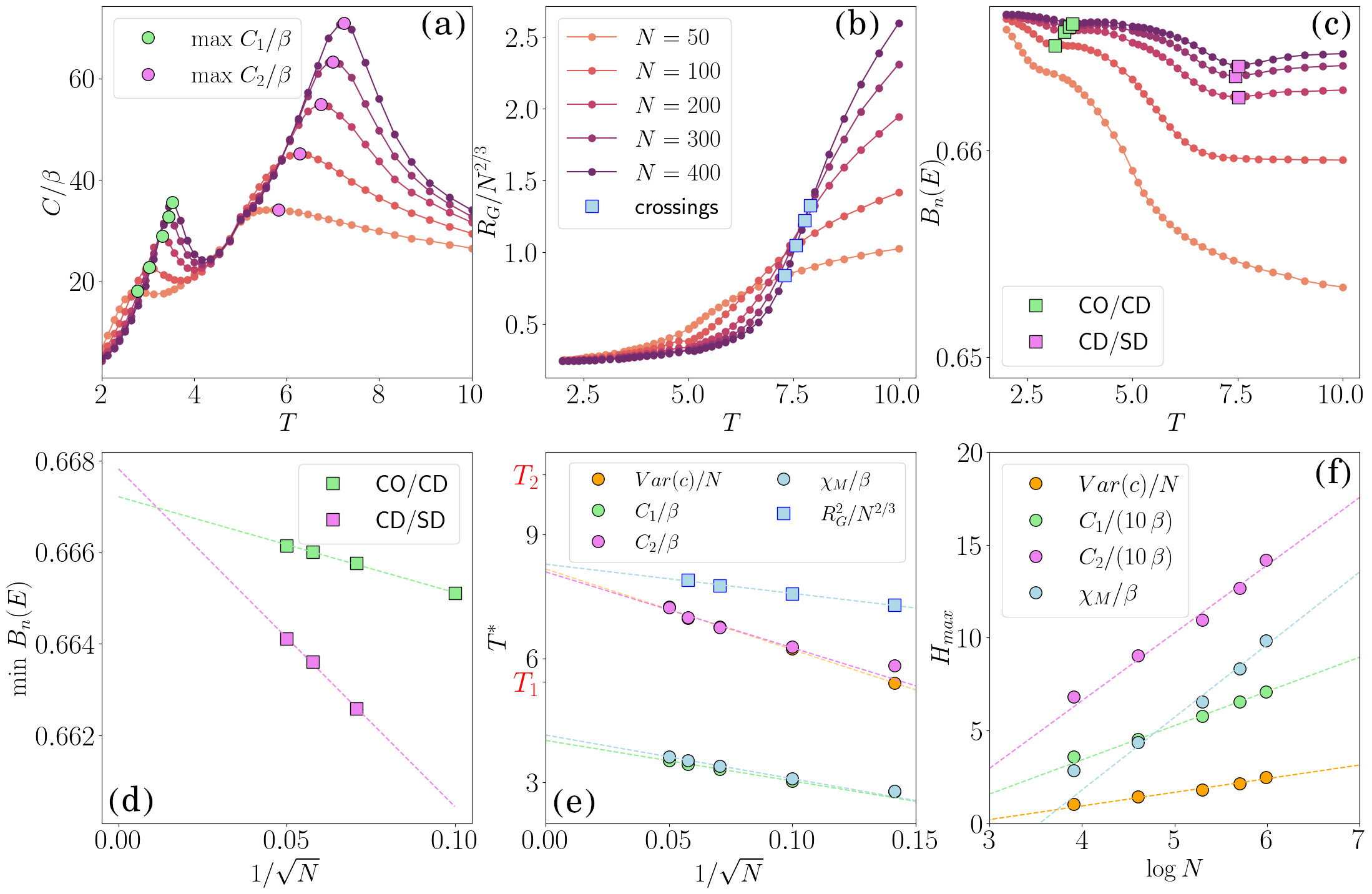}
    \caption{Monte Carlo results for $K/J = 3$, $\Delta = -5$. In panels (a)-(c), different curves refer to different values of $N$ (see legend in (b)). (a) Variance of the total energy $C/\beta$ as a function of $T$. The estimated locations of the maxima are highlighted by larger coloured circles.
    (b) $R_G^2/N^{2/3}$ as a function of $T$. Crossings with the $N = 400$ curve are highlighted by blue squares. (c) Binder cumulant of the energy $B_n(E)$ as a function of $T$. Minima are marked with green (CO/CD transition) and violet (CD/SD transition) squares. 
    (d) Values of $T$ at which $B_n(E)$ has a minimum as a function of $N^{-1/2}$. 
    (e) Values of $T$ at which different quantities (reported in the legend) display a maximum as a function of $1/\sqrt{N}$; they converge linearly to an asymptotic ($N\to\infty$) value. We extrapolate estimates for the transition temperatures and highlight the mean-field predictions $T_1$ and $T_2$ on the y-axis. 
    (f) Height of the maxima of different quantities (reported in the legend; note that, for $C/\beta$, we plot $H'_{max} = H_{max}/10$) 
    as a function of $\log{N}$. 
    }
    \label{superplotCK3Dm5}
\end{figure*}

More quantitatively, we looked at the $T$ dependence of the specific heat (see Fig.~\ref{superplotCK3Dm5}a): the presence of two consecutive phase transitions is suggested by %the formation of 
two sets of peaks %in the specific heat 
whose %distance in $T$ and 
heights increase with $N$. The fact that the set of peaks at low values of $T$ refers to the CO/CD transition is confirmed by the concomitant formation of peaks in the magnetic susceptibility (see Fig.~\ref{all_chi_var}a in the Appendix). Instead, the set of peaks at high values of $T$ corresponds to the CD/SD phase transition, since the variance of the number of contacts shows similar behavior in the same range of temperatures (see Figs.~\ref{all_chi_var}b in the Appendix). 

An observable that better describes the conformational properties of the polymer, but is not easily accessible through the mean-field calculations, is the statistical size of the chain, which we measured in terms of its mean-squared radius of gyration $\langle R_G^2\rangle$. Since in the two compact phases $\langle R_G^2\rangle \sim N^{2/d}$, in Figure~\ref{superplotCK3Dm5}b we report $\langle R_G^2\rangle/N^{2/3}$ as a function of $T$. It can be seen that, at large values of $T$, the scaled curves increase with $N$. This is expected since $\langle R_G^2\rangle \sim N^{2\nu_{SAW}}$ for polymers in the swollen phase, with $\nu_{SAW} \approx 0.5889 > 1/2$~\cite{clisby_PRL_2010}. On the other hand, at low values of $T$, where the polymer is in a compact (either disordered or ordered) phase, data for large polymers collapse onto a master curve. The finite $N$ crossover between the SD and the CD phase is revealed by the crossings of these curves, whose location can be used as a further estimate of the location of the SD/CD phase transition.% (see below). 

In addition, we determine the nature of the transitions by looking at the Binder cumulant of the energy $B_n(E)$ as a function of $T$~\cite{Bindercum} (see  Figure~\ref{superplotCK3Dm5}c). The onset of two sets of minima further confirms the presence of two phase transitions. Moreover, as $N$ increases, both sets seem to converge to the limiting value $2/3$, a clear indication that both transitions are continuous (see Figure~\ref{superplotCK3Dm5}d).\\ 
 
Finally, we provide estimates for the location of the phase boundaries via finite-size scaling (FSS). The plots reported in Figure~\ref{superplotCK3Dm5}e,f corroborate the continuous nature of the transition, not only for the location of the maxima of $C$, $\chi_M$ and $Var(c)$, but also for the value of the peaks for the same quantities, as well as the crossings of the $R_G$ curves. By averaging the set of extrapolated values over the different observables, we obtained  $T_c(SD/CD) = 8.2 \pm 0.1$ and  $T_c(CD/CO) = 4.08 \pm 0.08$. Note that both estimates are significantly smaller than the mean-field ones.\\ 
In summary, the MC findings for $J/K=3$ and $\Delta=-5$  corroborate the mean-field picture, namely the presence of three equilibrium phases with the 
SD/CD and CD/CO phase boundaries, both of which are continuous.

\subsection{Simulations for $K/J=3,\Delta=0$}

Increasing $\Delta$ favors neutral spins, or in other words dilution increases. At $\Delta=0$ the mean field theory predicts that there should once again be two phase transitions, an SD/CD and a CD/CO transition. Whilst the CD/CO transition remains continuous, the SD/CD transition is predicted to be discontinuous in this case (see Fig.~\ref{e1e2K3}).\\
\begin{figure*}
    \centering    \includegraphics[width=0.9\linewidth]{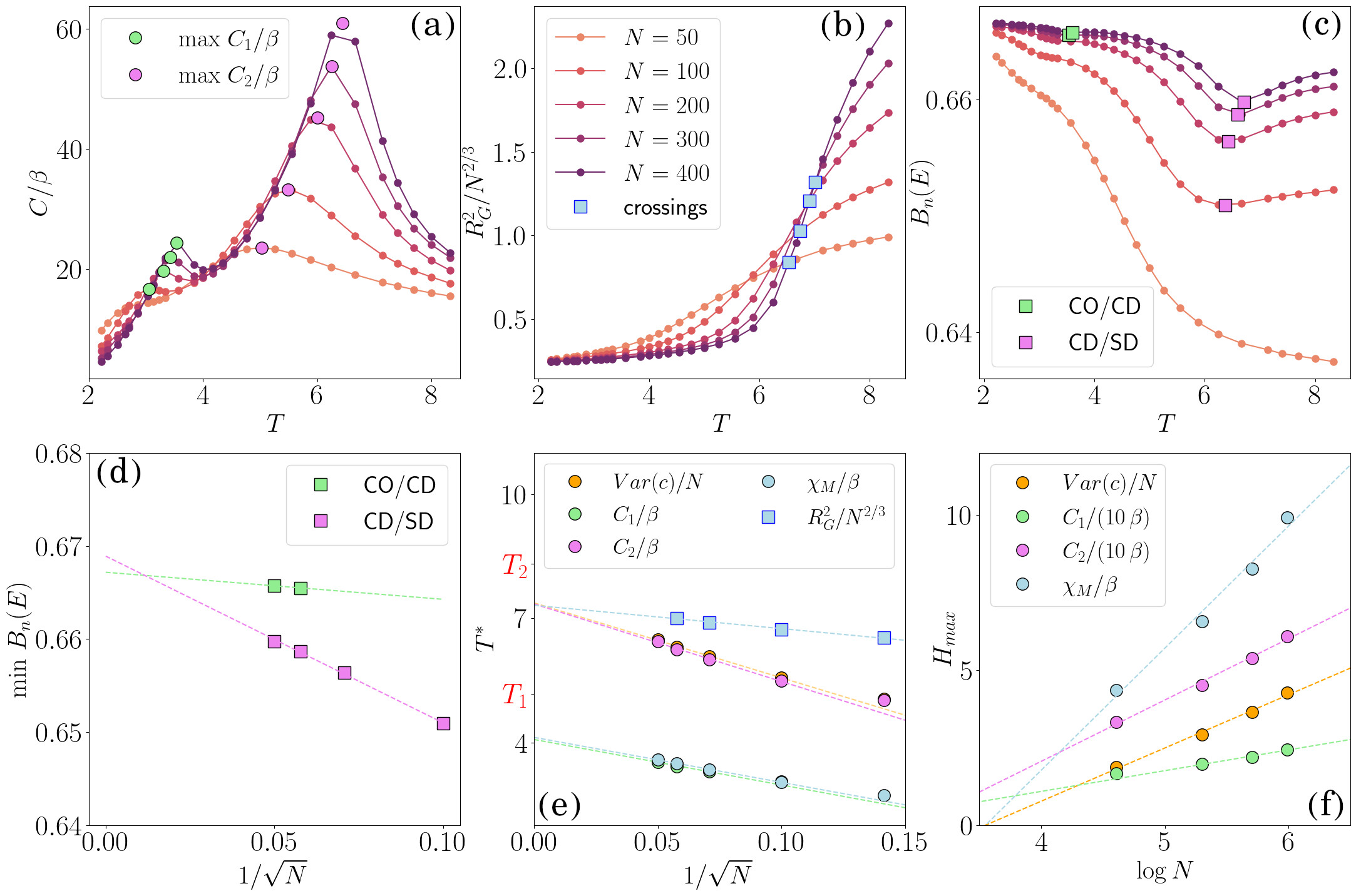}
    \caption{Monte Carlo results for $K/J = 3$, $\Delta = 0$. In panels (a)-(c), different curves and points are as in Fig.~\ref{superplotCK3Dm5}. (a) Variance of the total energy $C/\beta$ as a function of $T$. (b) $R_G^2/N^{2/3}$ as a function of $T$. 
    (c) Binder cumulant of the energy $B_n(E)$ as a function of $T$. (d) Values of $T$ at which $B_n(E)$ has a minimum as a function of $1/\sqrt{N}$. (e) Values of $T$ at which different quantities (reported in the legend) display a maximum as a function of $1/\sqrt{N}$. We extrapolate estimates for the transition temperatures and highlight the mean-field estimates $T_1$ and $T_2$ on the y-axis. 
    (f) Height of the maxima of different quantities (reported in the legend; for $C/\beta$, we plot $H'_{max} = H_{max}/10$) as a function of $\log{N}$.
    }
\label{superplotCK3D0}
\end{figure*}
The corresponding MC results indeed show two sets of peaks in the specific heat, corresponding to two transitions. The height of both sets of peaks increases with $N$ (see Fig.~\ref{superplotCK3D0}a). As for $\Delta=-5$, the set of peaks at low values of $T$ colocalises with the one observed in $\chi_M$ (see Figure~\ref{all_chi_var}c) and hence refers to the CO/CD phase transition. The second set, occurring at higher values of $T$,  originates from the corresponding non-monotonic behaviour of $Var(c)$ (see Figure~\ref{all_chi_var}d) and signals the SD/CD phase transition. This is confirmed by the crossings of the $R_G^2/N^{2/3}$ curves (see Fig.~\ref{superplotCK3D0}b). Notably, both sets of minima observed in the Binder parameter of the energy (see Fig.~\ref{superplotCK3D0}c) converge (as $N\to\infty$) to $2/3$, as expected for a continuous phase transition (see Figure~\ref{superplotCK3D0}d).

Finally, the linear trends of the location of the peaks of $C$, $\chi_M$, $Var(c)$, and the crossings of the scaled $R_g$ curves, once plotted as a function of $1/\sqrt{N}$ as well as the $\sqrt{N}$  growth of the peaks' heights (see  Figure~\ref{superplotCK3D0}e,f) corroborates this finding.

By averaging the extrapolated values of $T^*(N)$ on the different observables (see Figure~\ref{superplotCK3D0}), we estimate $T_c(SD/CD) = 7.3 \pm 0.1$ and $T_c(CD/CO) = 4.106 \pm 0.006$. Note that, again, the estimated values are lower than the mean-field ones.

Therefore, the Monte Carlo results at $\Delta=0$ confirm the mean-field prediction for the CO/CD phase boundary. However, there is a discrepancy regarding the nature of the SD/CD phase transition: while the mean field theory predicts a first-order transition, the magnetic polymer model in 3D is suggestive of a second-order transition.

To understand more in depth the relation between the mean field prediction and the behaviour of the system in 3D, we extended the MC investigation to $\Delta = 6$.

\subsection{Simulations for $K/J=3, \Delta=6$}
\begin{figure*}[t]
    \centering
    \includegraphics[width=0.9\linewidth]{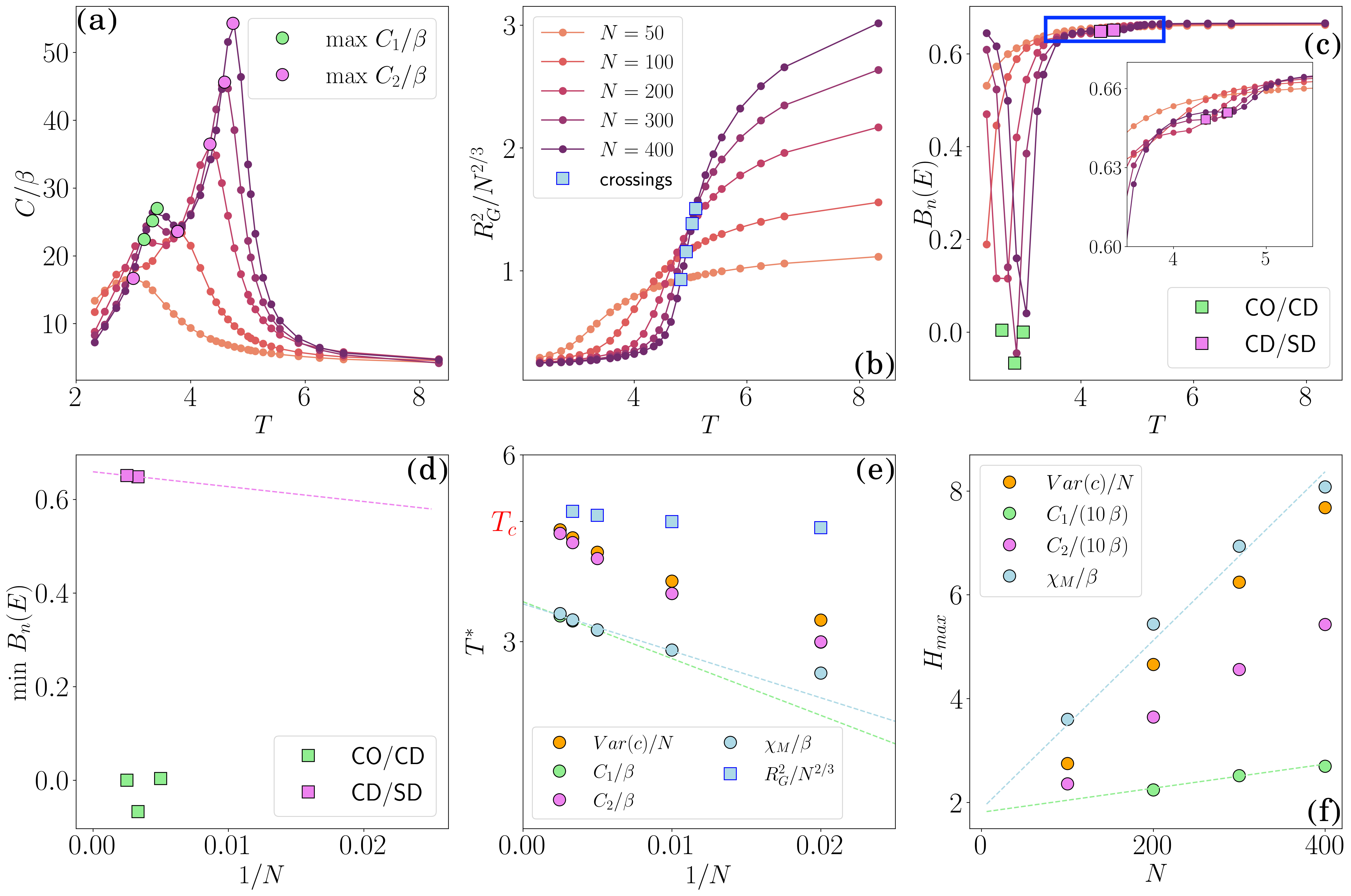}
\caption{Monte Carlo results for $K/J = 3$, $\Delta = 6$. In panels (a)-(c), different curves and points are as in Fig.~\ref{superplotCK3Dm5}. (a) Variance of the total energy as a function of $T$. (b) $R_G^2/N^{2/3}$ as a function of $T$. (c) Binder cumulant of the energy $B_n(E)$ as a function of $T$. (d) Values of $T$ at which $B_n(E)$ has a minimum as a function of $1/N$. (e) Values of $T$ at which different quantities (reported in the legend) display a maximum as a function of $1/{N}$. We extrapolate an estimate for the temperature corresponding to the CD/CO transition, highlighting the mean-field prediction $T_c$ on the y-axis. (f) Height of the maxima of different quantities (reported in the legend; for $C/\beta$, we plot $H'_{max} = H_{max}/10$) as a function of ${N}$.}
    \label{superplotCK3D6}
\end{figure*}
The mean-field theory predicts that, at $\Delta \approx 6.00$ and $T \approx 4.83$,
the system is at the multicritical point where the two first-order phase boundaries SD/SD and SD/CO meet the continuous phase transition line between CD and CO. According to what was observed at $\Delta=0$, we expect the 3D system to be in the region where the SD/CD and CD/CO boundaries are still well separated. We aim to verify whether the SD/CD boundary has become first-order as predicted by MF (see Figure~\ref{e1e2K3}a).

The onset of two distinct, although close, sets of peaks in the variance of the energy (see Fig.~\ref{superplotCK3D6}a) confirms the presence of two phase boundaries that, according to the concomitant developments of the peaks in $\chi_M/\beta$ and $Var(c)/N$, we can identify as the SD/CD and the CD/CO phase transitions, respectively (see Figs.~\ref{all_chi_var}e,f). The presence of the SD/CD transition at large values of $T$ is corroborated by the crossings of the radius of gyration curves when scaled by $N^{2/3}$, and by the minima of $B_n(E)$ (see Figures~\ref{superplotCK3D6}b,c).

We highlight that the set of minima of $B_n(E)$ at high values of $T$, which were already rather shallow at $\Delta=0$, are not observed in this case since all curves rapidly approach the value 2/3 in that range of $T$. This suggests that the CD/CO phase boundary remains continuous.

Instead, at low values of $T$, the set of minima converges as $N$ increases to a value that is far from $2/3$ (see Figure~\ref{superplotCK3D6}c,d): this is a clear signal that the CO/CD phase transition has now become first order.
We remark that the mean-field calculations predicted a qualitatively different scenario, with a continuous CO/CD transition and a first-order CD/SD transition. 
The discontinuous nature of the CO/CD transition is confirmed by the trends of the location of the peaks of $C/\beta$, and $\chi_M$, once plotted as a function of $1/N$, as well as the linear dependence in $N$ of the height of the peaks (see Figure~\ref{superplotCK3D6}e,f).
Thus, at $\Delta = 6$ the 3D system shows a change of the nature of a phase boundary from continuous to first-order; however, we observe it for the CO/CD boundary, rather than for the CD/SD one. This suggests that thermal fluctuations have a strong effect on the phase behaviour of the system and qualitatively modify the mean-field picture at high values of the dilution parameter $\Delta$, altering not only the location of the phase boundaries in the $(\Delta,T)$ plane but also their nature.

\section{Conclusions}

\begin{figure}
    \centering    \includegraphics[width=0.9\linewidth]{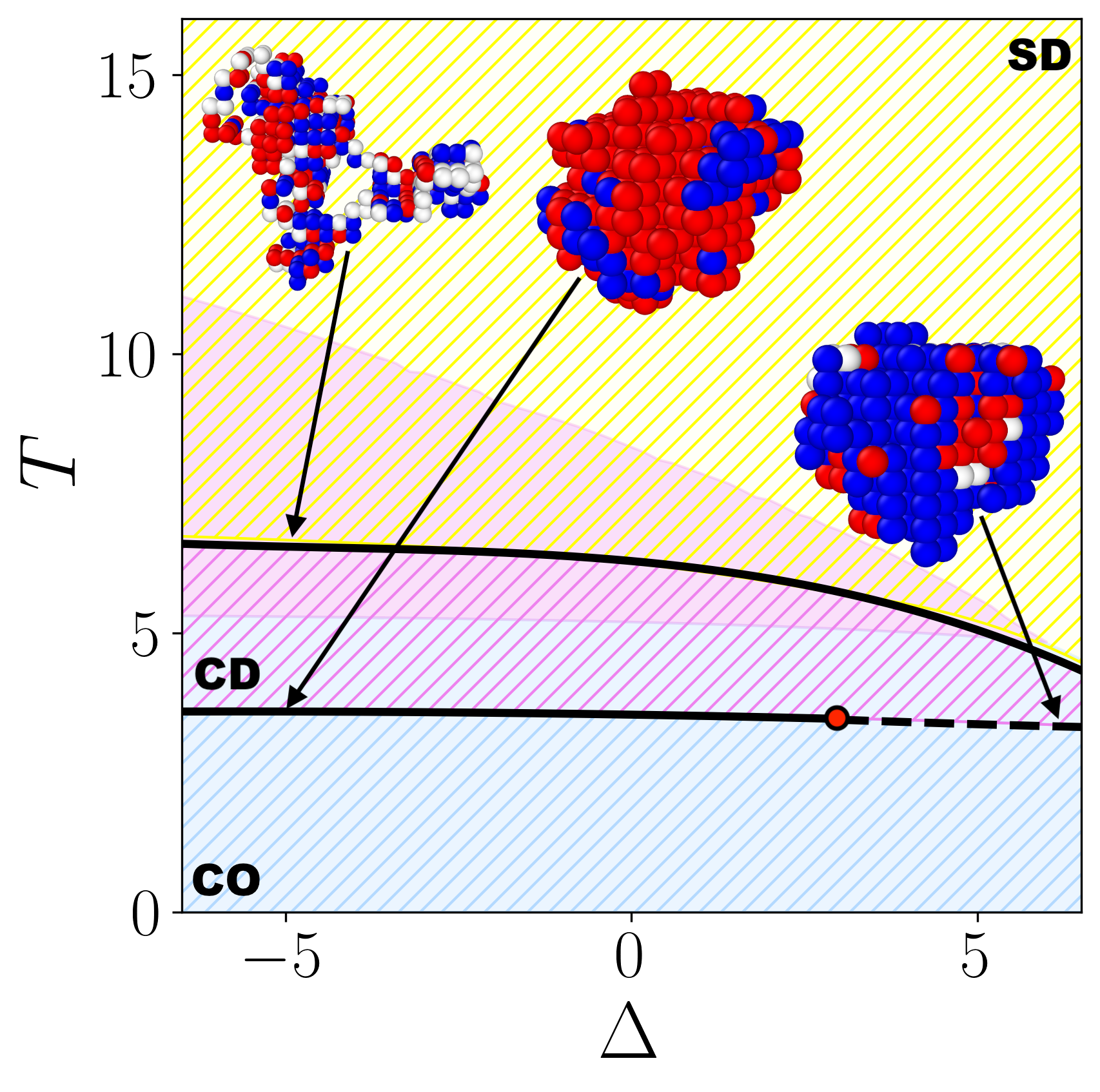}
    \caption{Estimated phase diagram of the magnetic polymer model on the cubic lattice ($K/J=3$) from Monte Carlo simulations.
    The diagonal yellow, magenta, and cyan patterns represent the regions with stable SD, CD, and CO phases respectively. These regions should be compared with the uniformly colored zones (same colors, lighter shade), which corresponds to the values of $(T,\Delta)$ at which the same phases are predicted to be stable by the mean-field (MF) theory.
    The solid and dashed lines represent respectively the continuous and discontinuous phase boundaries, as estimated by the Monte Carlo simulations; the red circle is the expected location of the multicritical point ($\Delta^*$, $T^*$).}
    \label{GreatFigureFinal}
\end{figure}

In this work, we have studied the equilibrium properties of a magnetic polymer endowed with a spin model (the Blume-Emery-Griffiths). There are two three terms in the Hamiltonian, which respectively describe: (i) exchange interactions between $\pm 1$ spins, favouring alignment (i.e., ferromagnetic interactions),  (ii) self-attraction between monomers and (iii) relative balance between neutral and $\pm 1$ spins, which we refer to as dilution (as increasing dilution increases the average portion of neutral spins on the chain). 

We first studied this problem using a mean-field (MF) approximation. 
The resulting MF phase diagram is very rich phase, and features phase boundaries whose nature and location depend on the degree of dilution $\Delta$ and the relative strength of the self-attraction parameter over the ferromagnetic one $K/J$.  
In addition to the previously reported swollen-disordered (SD) and compact-ordered (CO) phases~\cite{garel1999phase,papale2018ising,foster2021critical}, the MF results reveal the presence of an equilibrium phase characterized by a globular polymer conformation with zero magnetization.  This compact-disordered (CD) phase was not observed at equilibrium in previously studied models of magnetic polymers where the conformational transitions were triggered exclusively by the ferromagnetic coupling $J$.   

To validate these findings and to test the validity of the MF phase diagram against thermal fluctuations, we performed Monte Carlo simulations of the corresponding model on the cubic lattice $d=3$. We considered 
a fixed value of the $K/J$ ratio ($K/J=3$), at which the MF analysis displays a rich phase diagram as a function of the dilution parameter $\Delta$ (see Fig.~\ref{e1e2K3}). Our results show that, 
upon increasing $\Delta$, the MC picture fundamentally deviates from the mean-field one. At $\Delta=-5$, where dilution is weak, we confirm the presence and continuous nature of both phase transitions (CO/CD and CD/SD) predicted by the mean-field (MF) theory. 
At $\Delta = 0$, the nature of the CO/CD transition is confirmed and the transition temperature appears to be independent of $\Delta$ (as in mean-field). However, the CD/SD transition retains its continuous nature, at variance with the mean-field prediction (see Fig.~\ref{e1e2K3}). Upon increasing the dilution further ($\Delta=6$), we find that the scenario predicted by the MC simulations deviates even more from the mean-field, as the CO/CD transition becomes first-order, whereas the CD/SD remains second order -- this is the opposite of what MF predicts. Results and comparison are summarized in Figure~\ref{GreatFigureFinal}. 

Our results show that fluctuations are important in 3D, as they modify the nature of the transitions as predicted by the MF theory, most notably for the CO/CD transition. The traditional BEG model on the lattice (i.e., not on a polymer) also yields qualitative differences between mean-field calculations and Monte Carlo simulations have also been found on the lattice: for example, it has been shown that tricritical points may disappear, upon varying $K/J$, when moving from the microcanonical to the canonical ensemble \cite{hovhannisyan2017complete}. As such, the discrepancies between MC and MF reported here may align with what has been found for the traditional BEG model.

It is also of interest to discuss the results found here in the context of biophysics, where magnetic polymer models have previously been used to study the spreading of epigenetic marks on chromatin~\cite{michieletto2016polymer,michieletto2019nonequilibrium,coli2019magnetic,jost2018,owen2023}, the DNA-protein complex which provides the building block of eukaryotic chromosomes. With respect to those models, the main additional ingredient considered here is the $\pm 1$ spin dilution, or the presence of neutral spins, corresponding to inert or unmarked chromatin beads, devoid of active and inactive epigenetic marks (which can in turn be modeled as $\pm 1$ spins). This feature can effectively mimic the fact that marks need to be deposited by enzymes, which are limited in number in a living cell. Our results show that dilution leads to the emergence of the compact disordered phase, which was previously only found with polymer models out of thermodynamic equilibrium. The CD phase can represent chromatin regions compactified by bridging proteins which are not epigenetic readers or writers (an example of such a bridge could be cohesin or another SMC protein). Another important finding is that changing dilution can lead to a switch between a continuous and a discontinuous transition between the compact ordered and compact disordered phase. The order of the transition is relevant to the physics of chromatin, as a first-order transition endows the system with memory~\cite{michieletto2016polymer}, such that, for instance, it is now possible, within our model, for a compact and disordered state to retain its state following replication. 

%with their location shifted towards larger $\Delta$  and lower $T$ values . 
In the future it would be interesting to investigate numerically the $d=2$ case, where fluctuations can further affect the mean-field predictions, to see if the phase diagram changes also qualitatively. From the biophysics viewpoint, an avenue to extend our study would also be to introduce epigenetic bookmarks as in~\cite{michieletto2019nonequilibrium}, and investigate whether this can lead to the formation of stable unmarked inert, alongside marked active or inactive, epigenetic domains in chromatin.

\subsection*{Data availability statement}

The data supporting this article have been included as part of the Appendices.

\begin{acknowledgments}
E.O. acknowledges support from grant PRIN 2022R8YXMR funded by the Italian Ministry of University and Research. The authors acknowledge CloudVeneto for the use of computing and storage facilities.
\end{acknowledgments}

\section*{Appendix A: mean-field theory}

\subsection*{Hubbard-Stratonovich transformation}
To decouple the quadratic and bi-quadratic terms of the model Eq.~\eqref{partition}, we perform a double Hubbard-Stratonovich transformation and introduce two sets of local fields $\{\phi_i\}$ and $\{\alpha_i\}$. This gives
%\begin{widetext}
%\begin{equation}
\begin{align}
&\mathcal Z = \sum_{\gamma \in \text{SAW}} \sum_{\{S\}}  \int D\phi D\alpha \exp \left ( -\frac{1}{2 \beta J} \sum_{i,j}\phi_i (\Lambda_{i,j}^\gamma)^{-1} \phi_j +\right. \cr & - \frac{1}{2 \beta K} \sum_{i,j}\alpha_i (\Lambda_{i,j}^\gamma )^{-1}\alpha_j + \sum_{i} (\phi_i S_i  + \alpha_{i} S_{i}^{2} - \beta \Delta S_i^2) \cr &\,\, + \beta \Delta N \Bigg) 
\label{eq_HS1}
\end{align}
%\end{equation}
%\end{widetext}
 where $D\alpha$ and $D\phi$ are defined as:
 \begin{equation}
   D\phi \equiv  (2\pi)^{-N/2} (\beta J)^{-N/2} \det( \Lambda)^{-1/2} d^N \phi
 \end{equation}
  \begin{equation}
   D\alpha \equiv  (2\pi)^{-N/2} (\beta K)^{-N/2} \det( \Lambda)^{-1/2} d^N \alpha.
 \end{equation}
 By summing over all possible spin configurations $\{S_i\}$ we get  
%\begin{widetext}
\begin{align}
&\mathcal Z =  \sum_{\gamma \in \text{SAW}} \int D \phi D \alpha 
 \exp \left( -\frac{1}{2\beta J} \sum_{i,j} \phi_i (\Lambda_{{i,j}}^\gamma)^{-1} \phi_j \right. \cr    
& -\frac{1}{2\beta K} \sum_{i,j} \alpha_i (\Lambda_{{i,j}}^\gamma)^{-1} \alpha_j + \beta \Delta N \cr
&+\sum_{i = 1}^{N} \log  \left(1 + 2 e^{-\beta \Delta + \alpha_{i}} \cosh (\phi_{i} )\right)  \Bigg) 
\label{eq_HS2}
\end{align}
%\end{widetext}
The two integrals can be evaluated using a homogeneous saddle point approximation that replaces the integral with the maximum value of the integrand attained for $\phi_i=\phi$ and $\alpha_i=\alpha$.  This gives
%\begin{widetext}
\begin{align}
&\mathcal Z = A
\sum_{\gamma \in \text{SAW}}
\exp \left ( 
 -\frac{\phi^2}{2\beta J} \sum_{i,j}  (\Lambda_{{i,j}}^\gamma)^{-1}  
 -\frac{\alpha^2}{2\beta K} \sum_{i,j}  (\Lambda_{{i,j}}^\gamma)^{-1} \right. \cr
&+\sum_{i = 1}^{N} \log  (1 + 2 e^{-\beta \Delta + \alpha} \cosh (\phi )) + \beta \Delta N
\Bigg),
\label{eq:eq_Z3}
\end{align}
%\end{widetext}
where $A$ is a normalisation constant.
Since the term  $\sum_{i,j} \Lambda_{i,j}^\gamma$ depends on a given SAW it cannot be computed exactly. We can, however, approximate it by restricting the sum to the subset of space-filling SAWs within a volume $V$ known as Hamiltonian walks~\cite{garel1999phase}~\footnote{Hamiltonian walks are walks that visit each vertex of a lattice of volume $V$ exactly once and have been used as a lattice model of highly compact polymers}. For a given Hamiltonian walk, $\Lambda^{\gamma}$ simplifies to the adjacency matrix of the underlying lattice with coordination number $z$. 
This approximation is appropriate for SAWs in the compact phase, which is space-filling. However, it is rather strong for swollen SAWs: we mitigate this issue by replacing the lattice coordination number $z$ with  $\rho\,z$, thus considering SAWs that, having $\rho=N/V\,<1$, display an effective lower mean number of nearest neighbors. This gives
\begin{equation}
\sum_{i,j} \Lambda_{i,j}^\gamma \approx \frac{N}{z \rho}.
\label{Hamiltonian_walks_1}
\end{equation}
Within this approximation, the  number of N-steps SAWs confined in  a volume $V$ is given by~\cite{orland1985evaluation}
\begin{equation}
\mathcal Z_{SAW} = \left (\frac{z}{e} \right )^{N}\exp \left (-V (1 - \rho) \log (1 - \rho) \right ).
\label{Hamiltonian_walks_2}
\end{equation}

By plugging Eqs.~\eqref{Hamiltonian_walks_1} and ~\eqref{Hamiltonian_walks_2} in Eq.~\eqref{eq:eq_Z3} we obtain the  mean-field partition function as
%\begin{widetext}
\begin{align}
&\mathcal Z =  A \left (\frac{z}{e} \right )^{N}  \exp \left (-V (1 - \rho) \log (1 - \rho) -\frac{N \phi^2}{2 \beta z J \rho} \right. \cr  
& -\frac{N \alpha^2}{2 \beta z K \rho} + N \log (1 + 2 e^{-\beta \Delta + \alpha } \cosh (\phi + \beta h ))  \Bigg),
\label{part_MF}
\end{align}
%\end{widetext}
By taking the $-1/\beta \log$ of Eq. (\ref{part_MF}) and dividing the system's size $N$ we obtain the mean-field free energy density of Eq.\eqref{fed}.

\subsection*{Expansion of the mean-field free energy density }
Here we expand the free energy density Eq.~\eqref{fed} and the corresponding self-consistent equations Eqs.~\eqref{kmfe1},\eqref{kmfe2},\eqref{kmfe3} %
in the proximity of the different phase boundaries. % that can arise.

\subsubsection*{CO - SD phase boundary}
Since we expect a transition where $\phi$ and $\rho$ change from zero (SD) to a nonzero value (CO), we can expand around $\rho=0$ and $\phi=0$. 
By substituting the $\beta z \rho$ expression obtained from Eq.~\eqref{kmfe2} into Eq.~\eqref{kmfe1},  and then   Taylor expanding Eq.~\eqref{kmfe3} around $\rho = 0$, we obtain  
\begin{equation}
    \alpha(\phi) = \frac{K}{J}\phi \coth (\phi),
    \label{important1}
\end{equation}
that is Eq.~\eqref{alpha_vs_phi} of the main text, and
\begin{equation}
    \rho ^2 = \frac{\phi ^2}{ \beta J z}  + \frac{\alpha ^2}{\beta K z}.
    \label{important2}
\end{equation}
Expanding up to second order Eq.~\eqref{important1} around $\phi = 0$ and plugging the result in Eq.~\eqref{important2}, we find a simple expression for the local field $\phi$ as a function of $\rho$:
\begin{equation}
    \phi(\rho)  = \frac{\sqrt{\beta  J^2 \rho ^2 z-K}}{\sqrt{J}}.
    \label{phi_vs_rho}
\end{equation}
Finally,  by inserting Eq.~\eqref{phi_vs_rho} in Eq.~\eqref{fed}, Taylor expanding around $\rho = 0$ and computing the osmotic pressure as
\begin{equation}
    \Pi(\beta, \rho) = -\frac{\partial f}{\partial (1/\rho)} = \rho^2 \frac{\partial f}{\partial \rho},
    \label{osmotic}
\end{equation}
we obtain the virial expansion:
\begin{equation}
    \beta \Pi(\beta, \rho) = B_1(\beta, \rho)\rho + B_2(\beta, \rho)\rho^2 + B_3(\beta, \rho)\rho^3 + \dots
\end{equation}
with
\begin{subequations}
 \label{eq:cosd}
\begin{align}
B_{2}(\beta, \Delta) &=  1, \label{eq:cosd1}\\
B_{3}(\beta, \Delta) &= \frac{1}{3  } - \frac{2 \beta J^{3/2} \sqrt{K} z e^{K/J} \sin \left(\frac{\sqrt{K}}{\sqrt{J}}\right)}{2 K e^{K/J} \cos \left(\frac{\sqrt{K}}{\sqrt{J}}\right)+K e^{\beta  \Delta }}, \label{eq:cosd2} \\
B_{4}(\beta, \Delta) &= \frac{1}{4}. \label{eq:cosd3}
\end{align}
\end{subequations}
Note that the second virial coefficient is always positive, an indication that the SD-CO transition is always first order.\\ 

\subsubsection*{CD - SD phase boundary}
Since this transition involves two disordered phases, it is convenient to fix $\phi = 0$ in Eq.~\eqref{kmfe2}, solve it for $\alpha$, and then plug this expression into Eq.~\eqref{fed}. Finally, by computing the corresponding  osmotic pressure and expanding it for $\rho \simeq 0$ we obtain the virial coefficients
\begin{subequations}
\begin{align}
B_{2}(T, \Delta) &= 1 -\frac{2 \sqrt{\beta  K z}}{ \left(e^{\beta  \Delta }+2\right)}\\
B_{3}(T, \Delta) &= \frac{1}{3  }-\frac{2 \beta K z e^{\beta  \Delta }}{\left(e^{\beta  \Delta }+2\right)^2}\\
B_{4}(T, \Delta) &= \frac{1}{4 }-\frac{\beta K z e^{\beta  \Delta } \left(e^{\beta  \Delta }-2\right) \sqrt{\beta  K z}}{3 \left(e^{\beta  \Delta }+2\right)^3},
\end{align}
\end{subequations}
\subsubsection*{CD - CO phase boundary}
\begin{figure}[h!]
    \centering
    \includegraphics[width=0.45\textwidth]{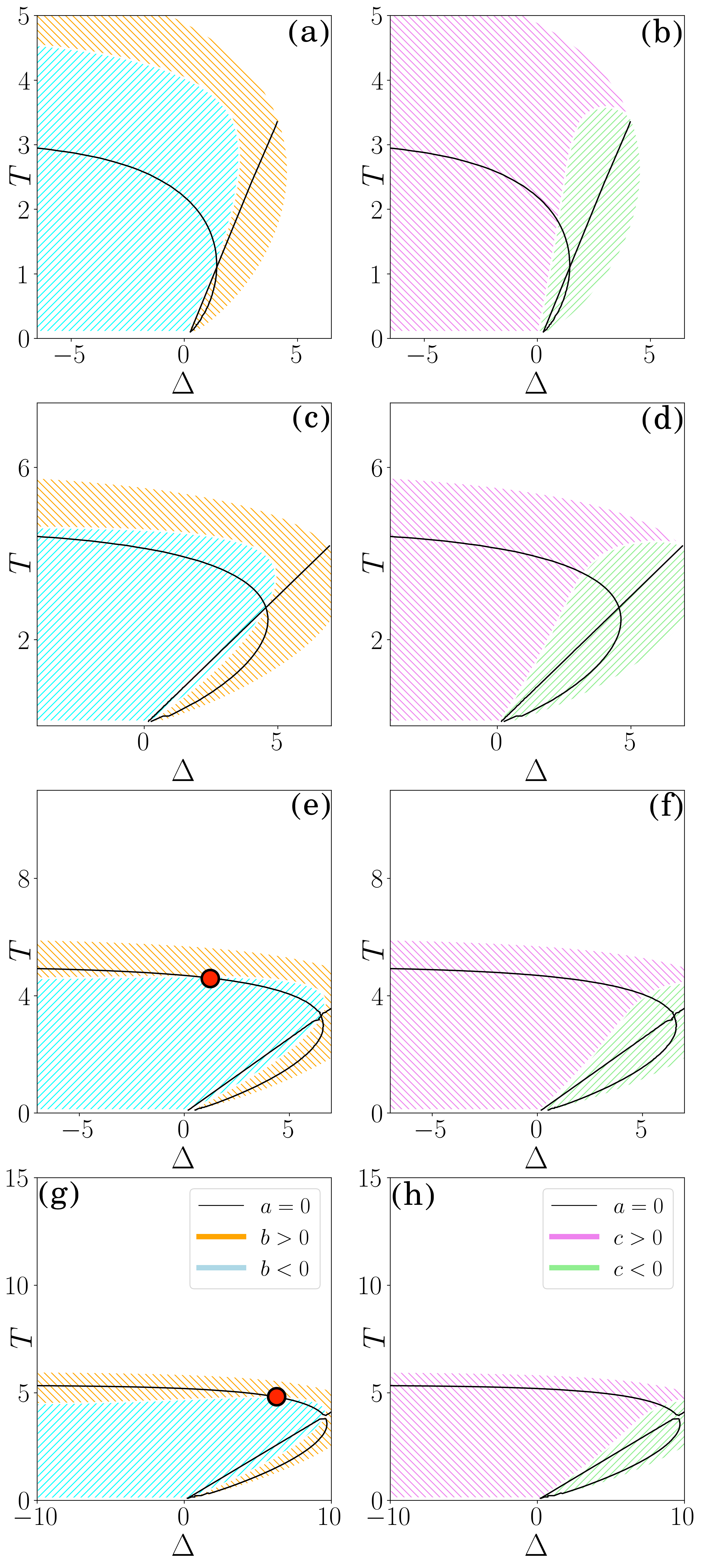}
    \caption{Contour plots of the coefficients $b(\Delta, T)$ and $c(\Delta , T)$ for $K/J = 0.8$ (a,b), $1.8$ (c,d), $2.3$ (e,f) and $3.0$ (g,h). In all panels, red dots mark the position of multicritical points, while black lines mark the values where $a(\Delta, T) = 0$. In the full white regions the functions $b(\Delta, T)$ and $c(\Delta, T)$ are not defined. Colored-white patterned regions (see legend in panels g,h) represent the regions where $b(\Delta, T)$ and $c(\Delta, T)$ have definite sign, separated by white lines marking either $b(\Delta, T)=0$ (left column) or $c(\Delta, T)=0$ (right column).}
    \label{fig:abc_CDCO}
\end{figure}
In this case the transition is between two compact phases ($\rho \ne 0$), one magnetically disordered and one magnetically ordered. We first plug the expression for $\alpha$ from Eq.~\eqref{important1} into Eq.~\eqref{kmfe1} and solve for $\rho$. This gives 
\begin{equation}
\rho(\phi) = \frac{\phi \, e ^ {\beta \Delta - K/J \phi \coth \phi}}{2 \beta J z \sinh \phi} + \frac{\phi \coth \phi}{\beta J z}
\label{rhod}
\end{equation}
Inserting Eqs.~\eqref{important1} and~\eqref{rhod} in Eq.~\eqref{fed} and expanding the result  around $\phi = 0$ we find the polynomial expression
\begin{equation}
    f(\beta,  \Delta) = 
f_0 + a(\beta,  \Delta)\phi ^2 + b(\beta,  \Delta)\phi ^4 + O(\phi ^6). 
\end{equation}
The explicit analytic expressions of $a(\beta,\Delta)$, $b(\beta,\Delta)$ and $c(\beta,\Delta)$ are cumbersome and will not be reported explicitly. %

In Fig.~\ref{fig:abc_CDCO} we show the contour plots of the coefficient $b(\beta,\Delta)$ and $c(\beta,\Delta)$ for the four different values of the ratio $K/J$ discussed in the main text. Condition $a(\Delta, T) = 0$ is represented on the plane as black lines. For all $K/J$ considered, this locus of points is made of straight and curved lines, originating from $\Delta = 0$ and $T = 0$. Condition $b(\Delta, T) = 0$ is displayed instead as a white line. Both this white line and the curved black line display a horizontal asymptote as $\Delta \to -\infty$. From Landau's theory of phase transitions, critical points arise when $a(\Delta, T) = 0$,  $b(\Delta, T) > 0$, and $c(\Delta, T) > 0$. Instead, a tricritical point is observed when $a(\Delta, T) = b(\Delta, T)= 0$ and $c(\Delta, T) > 0$. Thus, despite the presence of the straight black line, no critical points are observed in the green regions where $c(\Delta, T) < 0$ (see Fig.~\ref{fig:abc_CDCO}b,d,f,g). When $K/J = 0.8$ (Fig.~\ref{fig:abc_CDCO}a,b) the black line lies totally inside the blue region, where $b(\Delta, T) < 0$: no critical points appear on the CO/CD phase boundary. When $K/J = 1.8$ (Fig.~\ref{fig:abc_CDCO}c,d) the black line intercepts the white one when $\Delta \to -\infty$: thus a tricritical point is observed when there are no neutral states.
Indeed, we compute, numerically, the values of $K/J$ and $\beta$ such that a continuous CO/CD transition occurs: we solve the system of equations
\begin{equation}
    \left\{\begin{split}
    a_{\Delta \to -\infty}(\beta, K) = 0 \cr
    b_{\Delta \to -\infty}(\beta, K) = 0
    \end{split}\right.
    \label{system}
\end{equation}
where
%\begin{widetext}
\begin{align}
   a_{\Delta \to -\infty}(\beta, K)&= -\frac{2 \beta  J^2 z \log \left(1-\frac{1}{\beta  J z}\right)+2 J+K}{6 \beta  J} \\
   b_{\Delta \to -\infty}(\beta, K) &= \frac{24 \beta  J^2 z (\beta  J z-1) \log \left(1-\frac{1}{\beta  J z}\right)}{180 \beta  J (\beta  J z-1)} \cr & +\frac{9 \beta  J^2 z+2 \beta  J K z+J-2 K}{180 \beta  J (\beta  J z-1)}. 
  %b_{\Delta \to -\infty}(\beta, K) &= \frac{9 \beta  J^2 z+24 \beta  J^2 z (\beta  J z-1) \log \left(1-\frac{1}{\beta  J z}\right)+2 \beta  J K z+J-2 K}{180 \beta  J (\beta  J z-1)}.
\end{align}
%\end{widetext}
Such a system of equations cannot be solved analytically because of the logarithm appearing in both expressions. However, setting $J = 1$ and $z = 6$, we can solve Eq.~\eqref{system} numerically, to obtain $K \approx 1.8$ and $\beta \approx 0.2$. As such, for $K \ll 1.8$ the transition is always discontinuous, at $K \simeq 1.8$ becomes continuous at very negative values of $\Delta$, and for $K \gg 1.8$ the transition is always continuous, in agreement with the results reported in Figs.~\ref{c1c2}-\ref{e1e2K3}.

By increasing the $K/J$ ratio the tricritical point moves towards higher values of $\Delta$ and in Fig.~\ref{fig:abc_CDCO}e it is represented as a red dot. Increasing further the $K/J$ ratio, the tricritical point moves further towards the right (Fig.~\ref{fig:abc_CDCO}g), but in the phase diagram shown in Fig.~\ref{e1e2K3}a a CO/CD tricritical point is not observed because the presence of a first-order CO/SD phase transition.

\section*{Appendix B: Monte Carlo simulations}
\setcounter{subsection}{0}
\subsection*{Variance of magnetisation and contacts}
We report here the data regarding the dependence of the magnetic susceptibility per spin $\chi_M$, and of the variance of the number of contacts $Var(c)$ as a function of $T$ for the different systems considered ($\Delta=-5,0,6$).
\begin{figure}[]
    \centering
    \includegraphics[width=\linewidth]{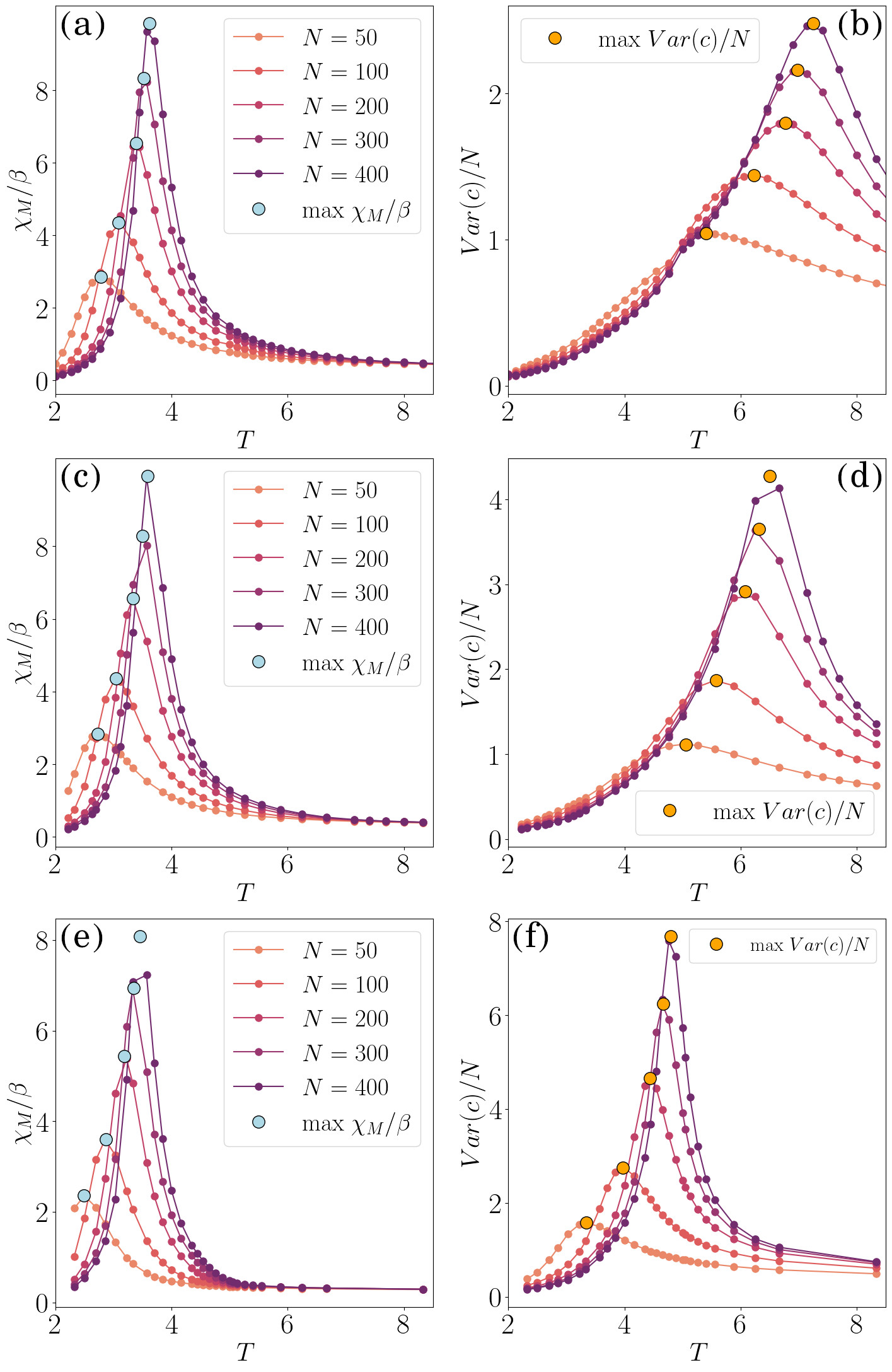}
    \caption{(a),(c),(e) magnetic susceptibility per spin $\chi_M/\beta$ and (b),(d),(f) variance of the number of contacts $Var(c)/N$ as a function of $T$ for the systems studied in the main text, $K/J = 3$ and (a),(b) $\Delta = -5$, (c),(d) $\Delta = 0$, (e),(f) $\Delta = 6$.}
    \label{all_chi_var}
\end{figure}
Figure~\ref{all_chi_var} reports such data for $\Delta=-5$ (panels a,b), $\Delta=0$ (panels c,d), and $\Delta=6$ (panels e,f). We observe that the sets of peaks for $\chi_M/\beta$ always occur at low temperatures and correspond, as highlighted in the main text, to the CO/CD transition. We note that the positions of these maxima do not change much with changing $\Delta$, in agreement with the mean-field predictions. Indeed, in Figure~\ref{e1e2K3}a the transition line between the CO and CD phases in the $\Delta$-$T$ plane is flat and practically does not depend on the temperature.\\Instead, the sets of peaks of $Var(c)/N$ always happen at high values of $T$ and correspond to the CD/SD transition. In this case, the positions of these maxima shift to lower values of $T$ with increasing $\Delta$, which is also in qualitative agreement with the mean-field predictions.

\subsection*{Correlation times and error bars}
Tables~\ref{tab1},~\ref{tab2} and \ref{tab3} report the estimates of the integrated autocorrelation times~\cite{madras1988pivot} for $\Delta = -5$, $\Delta = 0$ and $\Delta = 6$, respectively. Since data are collected every $50$ MC steps, the data reported use this quantity (that is, the sampling time) as the time unit.\\
\begin{figure}[]
    \centering
    \includegraphics[width=0.8\linewidth]{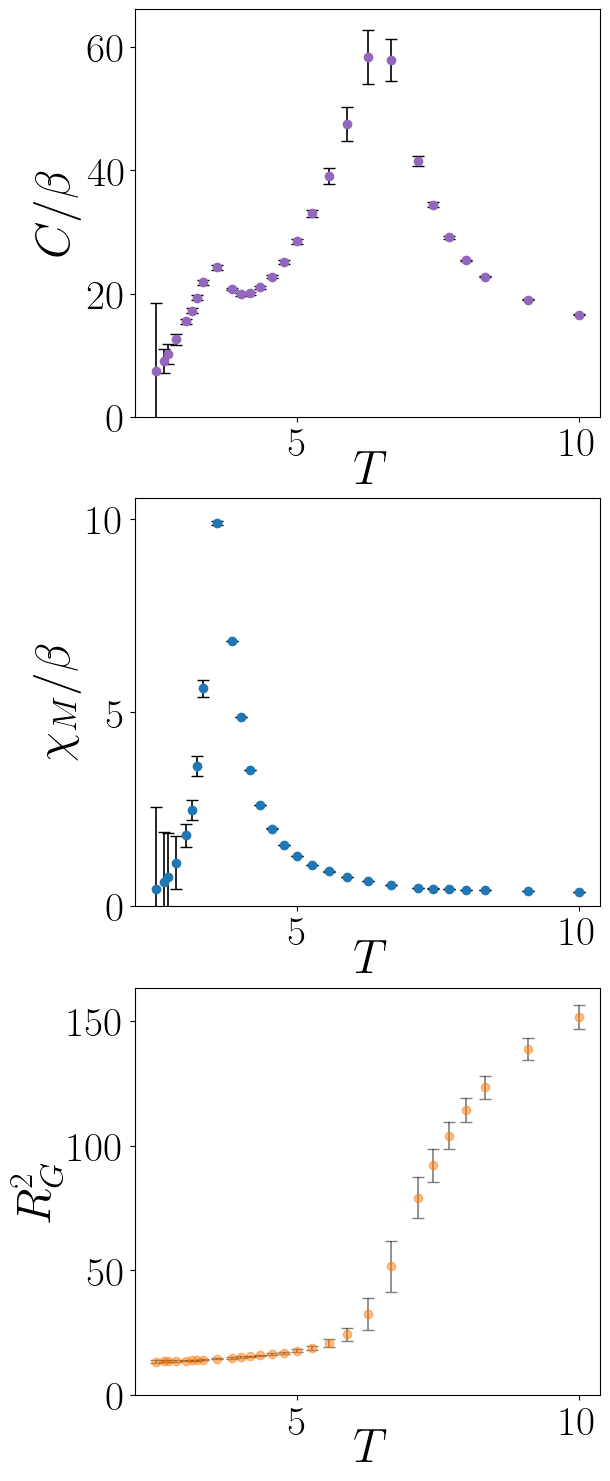}
    \caption{Estimations of (a) $C/\beta$, (b) $\chi_M/\beta$ and (c) $R_G^2$, complete with error bars, when $K = 3$, $J = 1$, $\Delta = 0$ and $N = 400$. %
    }
    \label{barreErrore}
\end{figure}
These estimates contribute to the error bars of the observables since the variance of an observable $O$ is  $2\tau_{O}$ larger than it would for statistically independent samples.
In Tables~\ref{tab1},~\ref{tab2},~\ref{tab3} we report the values of $\tau_{O}$  for the total energy, the magnetisation, and the radius of gyration.

The $T$ dependence of some observables with their corresponding error bars is exemplified in Figure~\ref{barreErrore}, where we reported the data at $\Delta=0$. We note that the largest error bars appear at low temperatures: this is because of the difficulty in efficiently sampling the system there, due to the large autocorrelation. Error bars are also large at temperatures close to the critical point, where critical slowing down is expected.
\begin{table}[h]
    \centering
    \begin{tabular}{|c|c|c|c|}
        \hline
        $T$ & $\tau_E$ & $\tau_M$ & $\tau_{RG}$ \\
        \hline
        $ 2.00 $ & $ 5.0(4) \cdot 10^4 $ & $ 2.0(2) \cdot 10^5 $ & $ 8.0(8) \cdot 10^4 $ \\
		$ 2.13 $ & $ 2.0(2) \cdot 10^4 $ & $ 1.2(5) \cdot 10^4 $ & $ 6.0(3) \cdot 10^3 $ \\
		$ 2.27 $ & $ 3.0(1) \cdot 10^3 $ & $ 5.0(2) \cdot 10^3 $ & $ 2.7(9) \cdot 10^3 $ \\
		$ 2.38 $ & $ 2.0(8) \cdot 10^3 $ & $ 2.7(8) \cdot 10^3 $ & $ 2.2(7) \cdot 10^3 $ \\
		$ 2.50 $ & $ 1.1(4) \cdot 10^3 $ & $ 4.5(7) \cdot 10^2 $ & $ 1.3(2) \cdot 10^3 $ \\
		$ 2.63 $ & $ 4.4(10) \cdot 10^2 $ & $ 2.3(2) \cdot 10^2 $ & $ 1.1(2) \cdot 10^3 $ \\
		$ 2.78 $ & $ 2.3(4) \cdot 10^2 $ & $ 2.3(2) \cdot 10^2 $ & $ 9.0(2) \cdot 10^2 $ \\
		$ 2.94 $ & $ 1.0(1) \cdot 10^2 $ & $ 1.0(8) \cdot 10^2 $ & $ 7.0(1) \cdot 10^2 $ \\
		$ 3.13 $ & $ 6.7(6) \cdot 10^1 $ & $ 3.9(2) \cdot 10^1 $ & $ 6.0(1) \cdot 10^2 $ \\
		$ 3.33 $ & $ 4.7(4) \cdot 10^1 $ & $ 1.03(3) \cdot 10^1 $ & $ 6.0(1) \cdot 10^2 $ \\
		$ 3.45 $ & $ 3.9(4) \cdot 10^1 $ & $ 4.92(8) \cdot 10^0 $ & $ 5.8(8) \cdot 10^2 $ \\
		$ 3.57 $ & $ 3.2(3) \cdot 10^1 $ & $ 2.39(3) \cdot 10^0 $ & $ 5.6(7) \cdot 10^2 $ \\
		$ 3.70 $ & $ 2.8(2) \cdot 10^1 $ & $ 1.202(8) \cdot 10^0 $ & $ 5.0(6) \cdot 10^2 $ \\
		$ 3.85 $ & $ 2.9(2) \cdot 10^1 $ & $ 7.62(3) \cdot 10^{-1} $ & $ 4.7(6) \cdot 10^2 $ \\
		$ 4.00 $ & $ 2.9(2) \cdot 10^1 $ & $ 5.9(2) \cdot 10^{-1} $ & $ 4.3(6) \cdot 10^2 $ \\
		$ 4.17 $ & $ 3.1(2) \cdot 10^1 $ & $ 5.29(1) \cdot 10^{-1} $ & $ 3.9(5) \cdot 10^2 $ \\
		$ 4.35 $ & $ 3.1(2) \cdot 10^1 $ & $ 5.1(1) \cdot 10^{-1} $ & $ 3.7(5) \cdot 10^2 $ \\
		$ 4.55 $ & $ 3.2(2) \cdot 10^1 $ & $ 5.02(8) \cdot 10^{-1} $ & $ 3.9(5) \cdot 10^2 $ \\
		$ 4.76 $ & $ 3.4(2) \cdot 10^1 $ & $ 5.01(8) \cdot 10^{-1} $ & $ 3.9(5) \cdot 10^2 $ \\
        $5.00$ & $1.7(8) \cdot 10^1$ & $5.01(7) \cdot 10^{-1}$ & $6.8(4) \cdot 10^1$ \\
        $5.13$ & $1.6(8) \cdot 10^1$ & $5.01(1) \cdot 10^{-1}$ & $6.3(4) \cdot 10^1$ \\
        $5.26$ & $1.72(9) \cdot 10^1$ & $5.01(7) \cdot 10^{-1}$ & $6.9(6) \cdot 10^1$ \\
        $5.41$ & $1.9(1) \cdot 10^1$ & $5.0(4) \cdot 10^{-1}$ & $7.0(5) \cdot 10^1$ \\
        $5.56$ & $2.1(1) \cdot 10^1$ & $5.0(4) \cdot 10^{-1}$ & $7.2(6) \cdot 10^1$ \\
        $5.71$ & $2.4(2) \cdot 10^1$ & $5.0(1) \cdot 10^{-1}$ & $7.5(6) \cdot 10^1$ \\
        $5.88$ & $3.0(2) \cdot 10^1$ & $5.0(4) \cdot 10^{-1}$ & $7.7(7) \cdot 10^1$ \\
        $6.06$ & $3.5(3) \cdot 10^1$ & $5.0(4) \cdot 10^{-1}$ & $8.0(7) \cdot 10^1$ \\
        $6.25$ & $4.0(3) \cdot 10^1$ & $5.01(1) \cdot 10^{-1}$ & $7.7(7) \cdot 10^1$ \\
        $6.45$ & $4.7(4) \cdot 10^1$ & $5.01(1) \cdot 10^{-1}$ & $6.7(5) \cdot 10^1$ \\
        $6.67$ & $5.6(4) \cdot 10^1$ & $5.02(1) \cdot 10^{-1}$ & $6.7(5) \cdot 10^1$ \\
        $6.90$ & $7.0(5) \cdot 10^1$ & $5.02(1) \cdot 10^{-1}$ & $6.8(5) \cdot 10^1$ \\
        $7.14$ & $6.6(5) \cdot 10^1$ & $5.0(4) \cdot 10^{-1}$ & $5.7(4) \cdot 10^1$ \\
        $7.41$ & $4.3(2) \cdot 10^1$ & $5.01(7) \cdot 10^{-1}$ & $3.3(2) \cdot 10^1$ \\
        $7.69$ & $2.7(1) \cdot 10^1$ & $5.0(4) \cdot 10^{-1}$ & $1.9(7) \cdot 10^1$ \\
        $8.00$ & $1.52(5) \cdot 10^1$ & $5.0(4) \cdot 10^{-1}$ & $1.0(3) \cdot 10^1$ \\
        $8.33$ & $7.8(2) \cdot 10^0$ & $5.0(4) \cdot 10^{-1}$ & $4.9(1) \cdot 10^0$ \\
        \hline
    \end{tabular}
    \caption{Autocorrelation times of the energy, magnetization and radius of gyration for $K/J = 3$, $\Delta = -5$ and $N = 400$. Data are reported using the sampling time ($M=$ 50 MC steps) as the unit of time.}
    \label{tab1}
\end{table}
\begin{table}
    \centering
    \begin{tabular}{|c|c|c|c|}
        \hline
        $T$ & $\tau_E$ & $\tau_M$ & $\tau_{RG}$ \\
        \hline
        2.22 & $1.6(5) \cdot 10^3$ & $2.0(2) \cdot 10^4$ & $6.0(4) \cdot 10^3$ \\
        2.33 & $6.0(2) \cdot 10^2$ & $7.0(4) \cdot 10^3$ & $2.0(1) \cdot 10^3$ \\
        2.50 & $2.2(4) \cdot 10^2$ & $2.2(8) \cdot 10^3$ & $6.0(1) \cdot 10^2$ \\
        2.63 & $1.0(2) \cdot 10^2$ & $9.0(2) \cdot 10^2$ & $3.4(7) \cdot 10^2$ \\
        2.70 & $5.7(6) \cdot 10^1$ & $6.0(2) \cdot 10^2$ & $3.6(7) \cdot 10^2$ \\
        2.86 & $3.7(4) \cdot 10^1$ & $2.4(5) \cdot 10^2$ & $2.1(4) \cdot 10^2$ \\
        3.03 & $1.5(1) \cdot 10^1$ & $6.6(8) \cdot 10^1$ & $3.3(4) \cdot 10^1$ \\
        3.13 & $1.17(7) \cdot 10^1$ & $3.3(2) \cdot 10^1$ & $1.5(2) \cdot 10^2$ \\
        3.23 & $1.02(6) \cdot 10^1$ & $2.2(1) \cdot 10^1$ & $2.3(2) \cdot 10^1$ \\
        3.33 & $9.3(5) \cdot 10^0$ & $1.31(5) \cdot 10^1$ & $3.6(4) \cdot 10^1$ \\
        3.57 & $6.8(3) \cdot 10^0$ & $2.94(4) \cdot 10^0$ & $3.3(3) \cdot 10^1$ \\
        3.85 & $5.3(2) \cdot 10^0$ & $7.82(5) \cdot 10^{-1}$ & $4.3(3) \cdot 10^1$ \\
        4.00 & $4.4(2) \cdot 10^0$ & $5.77(2) \cdot 10^{-1}$ & $3.5(2) \cdot 10^1$ \\
        4.17 & $4.2(2) \cdot 10^0$ & $5.28(3) \cdot 10^{-1}$ & $3.3(2) \cdot 10^1$ \\
        4.35 & $4.4(2) \cdot 10^0$ & $5.06(2) \cdot 10^{-1}$ & $2.3(2) \cdot 10^1$ \\
        4.55 & $4.5(2) \cdot 10^0$ & $5.01(1) \cdot 10^{-1}$ & $2.2(2) \cdot 10^1$ \\
        4.76 & $5.4(2) \cdot 10^0$ & $5.0(6) \cdot 10^{-1}$ & $1.7(1) \cdot 10^1$ \\
        5.00 & $6.5(3) \cdot 10^0$ & $5.0(6) \cdot 10^{-1}$ & $1.8(1) \cdot 10^1$ \\
        5.26 & $8.7(4) \cdot 10^0$ & $5.0(6) \cdot 10^{-1}$ & $2.4(1) \cdot 10^1$ \\
        5.56 & $1.52(1) \cdot 10^1$ & $5.0(6) \cdot 10^{-1}$ & $3.4(3) \cdot 10^1$ \\
        5.88 & $2.4(2) \cdot 10^1$ & $5.0(6) \cdot 10^{-1}$ & $3.7(2) \cdot 10^1$ \\
        6.25 & $3.6(2) \cdot 10^1$ & $5.0(6) \cdot 10^{-1}$ & $3.8(2) \cdot 10^1$ \\
        6.67 & $2.9(1) \cdot 10^1$ & $5.01(1) \cdot 10^{-1}$ & $2.4(1) \cdot 10^1$ \\
        7.14 & $9.5(3) \cdot 10^0$ & $5.03(2) \cdot 10^{-1}$ & $7.0(2) \cdot 10^0$ \\
        7.41 & $5.3(2) \cdot 10^0$ & $5.0(6) \cdot 10^{-1}$ & $3.7(1) \cdot 10^0$ \\
        7.69 & $3.3(1) \cdot 10^0$ & $5.0(6) \cdot 10^{-1}$ & $2.1(4) \cdot 10^0$ \\
        8.00 & $2.09(5) \cdot 10^0$ & $5.0(6) \cdot 10^{-1}$ & $1.51(3) \cdot 10^0$ \\
        8.33 & $1.78(4) \cdot 10^0$ & $5.0(6) \cdot 10^{-1}$ & $1.33(2) \cdot 10^0$ \\
        \hline
    \end{tabular}
    \caption{Autocorrelation times of the energy, magnetization and radius of gyration for $K/J = 3$, $\Delta = 0$ and $N = 400$. Data are reported using the sampling time ($M=$ 50 MC steps) as the unit of time.}
    \label{tab2}
\end{table}
\begin{table}[h]
    \centering
    \begin{tabular}{|c|c|c|c|}
        \hline
        $T$ & $\tau_E$ & $\tau_M$ & $\tau_{RG}$ \\
        \hline
        2.33 & $1.3(2) \cdot 10^2$ & $2.5(8) \cdot 10^3$ & $4.9(1) \cdot 10^2$ \\
        2.50 & $2.9(3) \cdot 10^1$ & $7.0(2) \cdot 10^2$ & $1.2(1) \cdot 10^2$ \\
        2.703 & $7.3(4) \cdot 10^0$ & $1.1(2) \cdot 10^2$ & $3.9(4) \cdot 10^1$ \\
        2.8 & $4.4(2) \cdot 10^0$ & $4.1(3) \cdot 10^1$ & $2.0(1) \cdot 10^1$ \\
        3.03 & $2.67(9) \cdot 10^0$ & $2.1(1) \cdot 10^1$ & $1.41(7) \cdot 10^1$ \\
        3.23 & $2.07(6) \cdot 10^0$ & $9.8(4) \cdot 10^0$ & $9.8(4) \cdot 10^0$ \\
        3.33 & $2.02(5) \cdot 10^0$ & $6.0(2) \cdot 10^0$ & $8.4(3) \cdot 10^0$ \\
        3.57 & $1.83(4) \cdot 10^0$ & $1.13(2) \cdot 10^0$ & $7.2(3) \cdot 10^0$ \\
        3.70 & $2.03(6) \cdot 10^0$ & $6.8(4) \cdot 10^{-1}$ & $6.4(2) \cdot 10^0$ \\
        3.85 & $2.35(8) \cdot 10^0$ & $5.5(3) \cdot 10^{-1}$ & $6.1(2) \cdot 10^0$ \\
        4.00 & $2.8(1) \cdot 10^0$ & $5.1(1) \cdot 10^{-1}$ & $5.9(2) \cdot 10^0$ \\
        4.17 & $3.08(9) \cdot 10^0$ & $5.06(2) \cdot 10^{-1}$ & $6.1(2) \cdot 10^0$ \\
        4.35 & $4.7(2) \cdot 10^0$ & $5.0(8) \cdot 10^{-1}$ & $6.2(2) \cdot 10^0$ \\
        4.44 & $5.7(2) \cdot 10^0$ & $5.0(8) \cdot 10^{-1}$ & $6.8(2) \cdot 10^0$ \\
        4.55 & $8.4(4) \cdot 10^0$ & $5.0(8) \cdot 10^{-1}$ & $8.0(3) \cdot 10^0$ \\
        4.65 & $1.1(6) \cdot 10^1$ & $5.0(8) \cdot 10^{-1}$ & $9.8(4) \cdot 10^0$ \\
        4.76 & $1.3(6) \cdot 10^1$ & $5.0(8) \cdot 10^{-1}$ & $1.1(5) \cdot 10^1$ \\
        4.88 & $1.25(6) \cdot 10^1$ & $5.0(8) \cdot 10^{-1}$ & $8.8(4) \cdot 10^0$ \\
        5.00 & $7.8(3) \cdot 10^0$ & $5.04(2) \cdot 10^{-1}$ & $5.4(2) \cdot 10^0$ \\
        5.05 & $6.8(3) \cdot 10^0$ & $5.0(2) \cdot 10^{-1}$ & $4.4(1) \cdot 10^0$ \\
        5.13 & $4.9(2) \cdot 10^0$ & $5.0(8) \cdot 10^{-1}$ & $3.3(1) \cdot 10^0$ \\
        5.26 & $3.17(1) \cdot 10^0$ & $5.0(8) \cdot 10^{-1}$ & $2.1(6) \cdot 10^0$ \\
        5.41 & $2.68(9) \cdot 10^0$ & $5.0(8) \cdot 10^{-1}$ & $1.55(3) \cdot 10^0$ \\
        5.56 & $2.14(7) \cdot 10^0$ & $5.01(1) \cdot 10^{-1}$ & $1.22(2) \cdot 10^0$ \\
        5.88 & $1.74(6) \cdot 10^0$ & $5.0(1) \cdot 10^{-1}$ & $9.9(1) \cdot 10^{-1}$ \\
        6.250& $1.38(4) \cdot 10^0$ & $5.0(8) \cdot 10^{-1}$ & $8.68(8) \cdot 10^{-1}$ \\
        6.67 & $1.06(2) \cdot 10^0$ & $5.0(8) \cdot 10^{-1}$ & $8.1(7) \cdot 10^{-1}$ \\
        8.33 & $9.2(2) \cdot 10^{-1}$ & $5.0(8) \cdot 10^{-1}$ & $7.2(7) \cdot 10^{-1}$ \\
        \hline
    \end{tabular}
    \caption{Autocorrelation times of the energy, magnetization, and radius of gyration for $K/J = 3$, $\Delta = 6$ and $N = 400$.  Data are reported using the sampling time ($M=$ 50 pivot moves) as the unit of time.}
    \label{tab3}
\end{table}

% Produces the bibliography via BibTeX.
\providecommand{\noopsort}[1]{}\providecommand{\singleletter}[1]{#1}%

\end{document}